\documentclass[journal]{IEEEtran}

 \usepackage{cite}
 \usepackage{amsmath,amssymb,amsfonts}
 \usepackage{algorithm}
 \usepackage{algorithmic}
  \usepackage{graphicx}
\usepackage{xcolor}
\usepackage{enumitem}
\usepackage{multirow}
\usepackage{booktabs,rotating}
\usepackage{flushend}
\usepackage{hyperref}
\usepackage{listings}
\usepackage{array}
\newcommand{\PreserveBackslash}[1]{\let\temp=\\#1\let\\=\temp}
\newcolumntype{C}[1]{>{\PreserveBackslash\centering}p{#1}}
\renewcommand{\algorithmicrequire}{\textbf{Input:}}
\renewcommand{\algorithmicensure}{\textbf{Output:}}
\usepackage{eqparbox}

\newcommand{\ie}{\textit{i.e.,~}}
\newcommand{\eg}{\textit{e.g.,~}}
\begin{document}

\title%[Multiple Targets Directed Greybox Fuzzing]
%{Kill Multiple Birds with One Stone: 
{Multiple Targets Directed Greybox Fuzzing}
%{Hunt Multiple Bugs at Once: Multiple Targets Directed Greybox Fuzzing}
%\author{Hongliang~Liang, \IEEEmembership{Member, IEEE,} Xianglin~Cheng, Jie~Liu, Jin~Li\\
\author{Hongliang~Liang, Xianglin~Cheng, Jie~Liu, Jin~Li\\
\thanks{\textit{H. Liang, X. Cheng, J. Liu are with Trusted Software and Intelligent System Lab, Beijing University of Posts and Telecommunications, Beijing, China. E-mail: \{hliang, chengxl, liujie\_ran\}@bupt.edu.cn.
J. Li is with Nation Key Laboratory of Science and Technology on Information System Security, Beijing, China. E-mail: tianyi198012@163.com.
This research is partially supported by CNKLSTISS.}}
}

\maketitle

%\IEEEauthorblockA{\textit{Trusted Software and Intelligent System Lab., Beijing University of Posts and Telecommunications, Beijing, China}}\\
%\IEEEauthorblockA{\textit{Nation Key Laboratory of Science and Technology on Information System Security (STISS)*}}
%\IEEEauthorblockA{\textit{School of computer science} \\
%\textit{Beijing University of Posts and Telecommunications}\\
%Beijing, China \\
%\{hliang, pyterware, xiaoda99, \}@bupt. edu. cn;~ tianyi198012@163.com
%}

\begin{abstract}Directed greybox fuzzing (DGF) can quickly discover or reproduce bugs in programs by seeking to reach a program location or explore some locations in order. However, due to their static stage division and coarse-grained energy scheduling, prior DGF tools perform poorly when facing multiple target locations (targets for short).

In this paper, we present multiple targets directed greybox fuzzing which aims to reach multiple programs locations in a fuzzing campaign. Specifically, we propose a novel strategy to adaptively coordinate exploration and exploitation stages, and a novel energy scheduling strategy by considering more relations between seeds and target locations. We implement our approaches in a tool called LeoFuzz and evaluate it on crash reproduction, true positives verification, and vulnerability exposure in real-world programs. Experimental results show that LeoFuzz outperforms six state-of-the-art fuzzers, \ie QYSM, AFLGo, Lolly, Berry, Beacon and WindRanger in terms of effectiveness and efficiency. Moreover, LeoFuzz has detected 23 new vulnerabilities in real-world programs, and 11 of them have been assigned CVE IDs.
\end{abstract}


%\newpage
\section{Introduction}

\textbf{Context.} Currently, fuzzing is one of the most effective and practical techniques to discover bugs or vulnerabilities automatically. By constantly mutating seeds initially provided, fuzzers generate lots of new inputs and report those that cause the program under test (PUT) failure or crash~\cite{fuzz}. A greybox fuzzer such as AFL~\cite{afl} uses program feedback like branch coverage to boost the efficiency of finding bugs. However, its consideration to achieve maximum code coverage may waste a lot of resources in some bug independent code.

By contrast, directed greybox fuzzers, \eg AFLGo~\cite{AFLGo}, Lolly~\cite{Lolly}, Berry~\cite{Berry}, spend most of the time budget 
in reaching target program locations (targets for short), \eg problematic changes, critical APIs or potential bugs, and thus are more suitable for patch testing and crash reproduction etc. For example, AFLGo uses harmonic distance between a seed and targets to reach the targets fast. Lolly exploits target statement sequences to trigger bugs which are resulted from the sequential execution of multiple statements. Berry uses concolic execution to enhance the directedness when reaching deep targets along some complex paths.

It is reasonable and meaningful for DGF to seek to reach multiple targets because there are often multiple bugs in real-world programs. To demonstrate the situation, we randomly selected nine widely-used programs in the real world and counted the bugs or vulnerabilities in them. As shown in Table~\ref{tab-mulbugs}, at least three CVEs were discovered in each of them.
Moreover, to expose or verify multiple (\eg $n$) bugs in a program via directed greybox fuzzing, one way is to run in parallel $n$ fuzzing instances, each of which is given a single target to trigger a single bug; another way is to run a directed greybox fuzzing instance with $n$ targets to trigger $n$ bugs, \eg AFLGo, Lolly and Berry.
% Moreover, to expose multiple (\eg $n$) bugs, one way is to run in parallel $n$ fuzzing instances each of which aims to trigger a single bug, \eg ClusterFuzz \cite{clusterfuzz} and PAFL \cite{pafl}; another way aims to cover $n$ targets in one directed fuzzing instance \emph{simultaneously}, \eg AFLGo, Lolly and Berry.
Though both methods are complementary, in this paper, our goal is to improve the effectiveness and efficiency of the second way on real-world programs.

\textbf{Problems.} Although DGF is efficient by spending more resources to explore the code towards the target locations, prior DGF tools~\cite{AFLGo,Lolly,Berry} perform poorly when facing multiple target locations due to their \emph{coarse-grained energy scheduling} and \emph{static stage division}.

% Fuzzers usually design an energy scheduling strategy to control the number of seed mutations.

An energy scheduling strategy is usually designed in the Fuzzers to control the number of seed mutations. In DGF, the scheduling strategy gradually adds (reduces) energy to seeds closer to (far away) target locations, which helps trigger multiple targets faster. For example, AFLGo gives more energy to a seed with a smaller harmonic distance to all targets, though this strategy makes AFLGo ignore the local optima.

\emph{Problem 1: }To cover multiple targets, pursuing a global optimal scheduling for all targets would ignore local optimal scheduling for some targets, as AFLGo does, while seeking an optimal scheduling for a single target would make other targets difficult to be reached, as Lolly and Berry do. Designing a suitable energy scheduling for reaching as many targets as possible within a time limit is another problem.

DGF works in two stages, \ie exploration and exploitation. In the exploration stage, the fuzzer aims to obtain more code coverage through seeds mutation and execution, thereby obtaining more run-time information. In exploitation stage, the fuzzer mutates and executes seeds to get seeds closer to the target locations. For instance, AFLGo begins with the exploration stage  first and randomly mutates the initial seeds to generate many new inputs,
in order to increase code coverage. It then enters exploitation stage to generate more new inputs which are increasingly closer to the targets. However, the time to enter exploitation stage is specified \emph{statically}. For example, AFLGo specifies 20 hours for exploration and 4 hours for exploitation. This static switching strategy ignores the dynamic runtime information and may decrease the performance of AFLGo.

\emph{Problem 2: }Less exploration would provide less coverage information for exploitation, making it difficult to generate high-quality directed seeds in exploitation stage. However, overmuch exploration would cost many resources and delay the exploitation, resulting in loss of directedness. Therefore, it is a challenge to coordinate exploration and exploitation stages in order to balance the coverage and directedness in DGF.

\begin{table}[t]
\renewcommand\arraystretch{1.2}
    \centering
    \caption{A real-world program usually contains multiple bugs.}
    \label{tab-mulbugs}
    \begin{tabular}{l|l|r}
    \hline
        Program & Version & \#Bugs \\ \hline
        cxxfilt & 2.26 & ~6~~\cite{cxxfilt} \\ 
        httpd & 2.4.46 & ~7~\cite{b10} \\ 
        jasper & 2.0.14 & 14~~\cite{b5} \\ 
        jasper & 2.0.12 & 15~~\cite{b6} \\ 
        libming & 0.4.8 & 70~~\cite{b7} \\ 
        objdump & 2.34 & ~4~~\cite{b1} \\ 
        readelf & 2.28 & ~8~~\cite{b2} \\ 
        sqlite & 3.32.0 & 10~~\cite{b8} \\ 
        tcpdump & 4.9.3 & 26~\cite{b9} \\ 
        tiff2pdf & 4.09 & ~3~~\cite{b3} \\ 
        tiff2pdf & 4.08 & ~4~~\cite{b4} \\ \hline
    \end{tabular}
\end{table}

\textbf{Proposal.} To solve the above problems, we present multiple targets directed greybox fuzzing to efficiently cover multiple programs locations in a single fuzzing campaign. Specifically, we propose a novel energy scheduling strategy that considers multiple relations between a seed and target sequences (MES for short) and a novel approach to adaptively coordinate exploration and exploitation stages (CEE for short) based on two queues.

As for energy scheduling, unlike AFLGo, which considers the harmonic distance between a seed and multiple targets in energy scheduling, and also unlike Lolly, which uses seed's target sequence coverage as the feature of energy scheduling, for a seed $s$ and multiple target sequences, MES first selects the target sequence $ts$ which has the highest coverage over $s$' execution trace, and then considers three relations between the seed and target sequences for energy scheduling, namely $s$'s sequence coverage ($seqCov$), $ts$' priority ($priority$) and global maximum coverage ($gMaxCov$). Specifically, MES assigns more energy to seeds with high $seqCov$, high $priority$, low $gMaxCov$, and vice versa. In this way, MES enables our fuzzer to reach as many targets as possible.

CEE uses a queue to store seeds which help to reach the targets (directed seeds for short) and another queue to store seeds that increase the code coverage (coverage seeds for short). If the proportion of the coverage seeds in the total seeds is too high (\eg exceeds a threshold $rate$) when exploring, our fuzzer switches to exploitation stage. If code coverage information is insufficient (\eg the fuzzer does not generate new directed seeds during a long period) when exploiting, our fuzzer turns to the exploration stage. Moreover, CEE adjusts the threshold $rate$ dynamically by recording the duration time and the number of generated directed seeds in each exploitation stage, in order to coordinate the exploration and exploitation stage adaptively.

\textbf{Evaluation.} We implemented the above techniques in a tool named LeoFuzz and conducted extensive experiments with seven real-world programs. Evaluation results demonstrate that LeoFuzz is effective and efficient on crash reproduction, true positives verification, and vulnerability discovery, compared to six state-of-the-art fuzzers \ie QSYM, AFLGo, Lolly, Berry, Beacon and WindRanger. Contrary to intuition, running a fuzzer with multiple targets in a single fuzzing campaign is more efficient than running multiple fuzzer instances in parallel with a target per instance.  In addition, LeoFuzz found in 
three real-world programs 23 new vulnerabilities and 11 of which are assigned CVE IDs.

\textbf{Contributions.} The main contributions of this paper are as follows:
\begin{itemize}
 \item An adaptive stage coordination approach which steers the fuzzer to switch between exploration stage and exploitation stage dynamically;

\item  A novel energy scheduling strategy which considers more relations between a seed and targets and hence enables to reach multiple targets efficiently;

\item A tool named LeoFuzz can expose and verify vulnerabilities in real-world programs. We make LeoFuzz publicly available\footnote{https://github.com/hongliangliang/leofuzz} to foster further research in the area;

\item  Extensive evaluation results show that LeoFuzz outperforms six state-of-the-art fuzzers, \ie QSYM, AFLGo, Lolly, Berry, Beacon and WindRanger, on crash reproduction and true positives verification. Moreover, LeoFuzz found in three real-world programs 23 new vulnerabilities and 11 of which are assigned CVE IDs.

\end{itemize}

The rest of this article is structured as follows. Our motivation is described in section \ref{motivation}. We present LeoFuzz's overall design in section \ref{approach},  static analysis  in section \ref{static-analysis},  dynamic analysis in section \ref{dynamic-analysis}, and its implementation in section \ref{Implement}. Section \ref{evaluation} presents the evaluation of LeoFuzz. We discuss the related work in section \ref{relwork},  threats to validity in section \ref{threats} and conclude in section \ref{conclusion}.

% We present LeoFuzz's design and implementation in section \ref{approach} and section \ref{Implement}. Section \ref{evaluation} presents the evaluation of LeoFuzz. We discuss the related work in section \ref{relwork} and conclude in section \ref{conclusion}.

\section{Motivation}\label{motivation}

In this section, we use an example to discuss two limitations of the existing DGF tools and introduce our approach.

\begin{figure}[t]
\centering
\includegraphics[height=6cm,width=6cm]{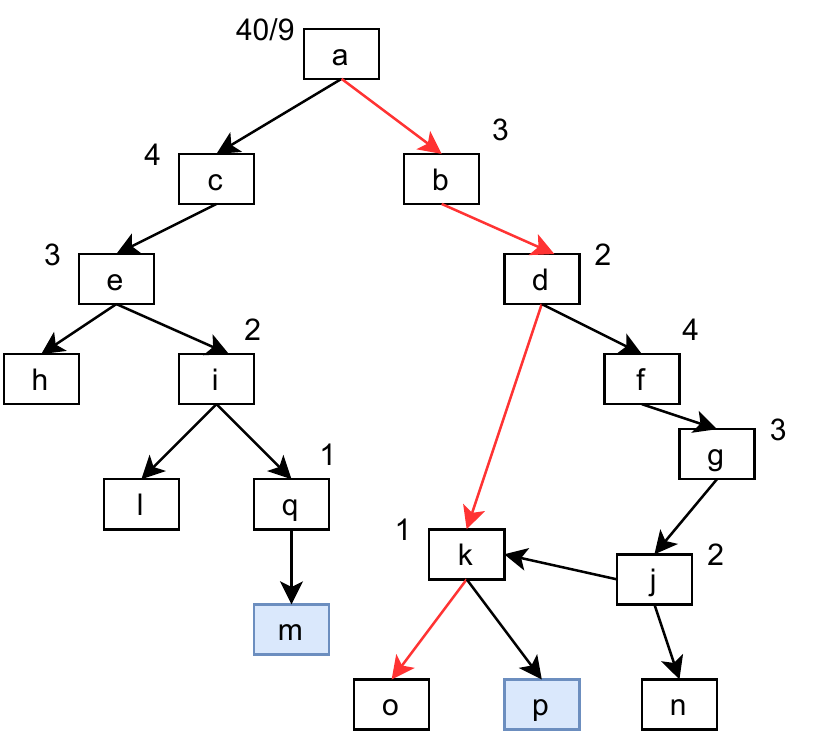}
\caption{A part of ICFG of objdump}
\label{icfg}
\end{figure}

Fig.~\ref{icfg} shows a part of the inter-procedural control flow graph (ICFG) of \texttt{objdump} program (V2.31). Each node in the figure indicates a basic block, whose details are shown in Table~\ref{table-example}. Two shadowed nodes, \ie $m$ and $p$, refer to an out-of-memory vulnerability and an integer overflow vulnerability respectively, \ie CVE-2018-13033 and CVE-2018-20671.
We ran DGF tools, \ie AFLGo, Lolly and Berry, with the locations of $m$ and $p$ as targets to trigger these bugs, and found two problems in these tools.

\begin{table}[t]
\renewcommand\arraystretch{1.2}
% \footnotesize
    \centering
    \caption{The program components for each node in Fig.\ref{icfg}}
    \label{table-example}
    \begin{tabular}{l|l|l|l}
    \hline
        Node & File & Line & Function \\ \hline
        a & objdump.c & 3682 & display\_object\_bfd  \\ 
        b & format.c & 234 & bfd\_check\_format\_matches \\ 
        c & objdump.c & 3688 & display\_object\_bfd  \\ 
        \multicolumn{4}{c}{......}  \\ 
        m & objdump.c & 2508 & load\_specific\_debug\_section  \\ 
        p & libbfd.c & 271 & bfd\_malloc  \\ \hline
    \end{tabular}
\end{table}

\textbf{Problem 1: unsuitable energy scheduling hinders covering multiple targets.} Existing DGF tools use two energy scheduling strategies. AFLGo adapts Dijkstra algorithm to schedule seeds, and Lolly or Berry exploits target sequence coverage in energy scheduling. When testing the \texttt{objdump} program, AFLGo had an execution trace for each of three seeds, \ie \texttt{a-c-e-i-q-m}, \texttt{a-b-d-k-o} and \texttt{a-b-d-f-g-j-k-p}, respectively. It calculates the harmonic distance ($d$, we label it aside the node in Fig. \ref{icfg}) between each node in these paths to two targets. For example, the harmonic distance of node $a$ is $d_a = 2 / (1/5 + 1/4) = 40/9$, where 5 and 4 is the length of the shortest path from $a$ to target $m$ and $p$ respectively, and 2 represents two targets that $a$ can reach. So the global distances of three seeds are $d_{aceiqm} = (40/9 + 4 + 3 + 2 + 1) / 5 = 2.89$,  $d_{abdko} = (40/9 + 3 + 2 + 1) / 4 = 2.61$, $d_{abdfgjkp} = (40/9 + 3 + 2 + 4 + 3 + 2 + 1) / 7 = 2.78$, respectively. AFLGo always selects the seed with the smallest global distance, \ie \texttt{a-b-d-k-o} here, though it is not reasonable as the seed covers none of two targets. In fact, the other two seeds, \ie \texttt{a-c-e-i-q-m} and \texttt{a-b-d-f-g-j-k-p}, reach the target $m$ and $p$ respectively, each of which should be selected. Therefore, AFLGo would ignore the local optima when seeking global optimal scheduling, thus reducing the directedness of fuzzing.

By contrast, Lolly or Berry may fall into an easy local optimum and thus never explore other deep targets. For example, if the paths going through the target $m$ are more difficult to explore (\eg due to complex conditions) than those going through the target $p$, the target sequence coverage of a seed close to $m$ would increase more slowly than that of another seed close to $p$. So both tools would generate a large number of seeds exploring the right branch of Fig. \ref{icfg}, and only a few seeds covering the left branch of Fig. \ref{icfg}. They would continue to explore the right branch of  Fig. \ref{icfg} even after reaching the target $p$. As a result, it is difficult for them to schedule a seed close to the target $m$. %to generate seeds to cover left branch code of Fig. \ref{icfg}, or to schedule a seed close to the target $m$.% even if  it is generated, it is difficult to be scheduled. their seed queues will have 
% So both tools would continue to explore the right branch of Fig. \ref{icfg} even after target $p$ is reached. As a result, it is difficult or impossible to schedule the seeds close to target $m$.

\textbf{Problem 2: improper exploration-exploitation division.}
Existing DGF tools switch between exploration and exploitation stage in three ways. The first one uses the seed selection strategy without considering the exploration-exploitation switchover, like Lolly~\cite{Lolly}. The generated coverage seeds or directed seeds are placed at the end of a queue for sequential scheduling. This method is simple but may take a long time to mutate those directed seeds at (or near) the rear of the queue, which slows down the reaching of targets. The second is the static division strategy used by AFLGo~\cite{AFLGo} and RDFuzz~\cite{RDFuzz}, which gives each stage a fixed period. This strategy is not flexible and does not consider the runtime information at all. The third is the exploitation-first strategy used in Berry~\cite{Berry}. It divides the seed queue into three priority levels, and directed seeds have higher priority than coverage seeds. This strategy doesn't work well when the fuzzer has insufficient code coverage information, hence causing a lower quality of the generated seeds.

\textbf{Our approach.} To solve the above problems, we propose and implement two techniques in LeoFuzz. 1) a fine-grained energy scheduling strategy, which considers more relations between seeds and targets, \eg target sequence priority, target sequence coverage and global maximum coverage. Our strategy can avoid ignoring the local optimum like AFLGo and avoid falling into an easy local optimum like Lolly or Berry. For instance, when seeds always have a higher coverage with the target sequence of $p$ than that of $m$, we can know that target $m$ is more difficult to reach than target $p$. Thus the global maximum coverage of the target $m$ is lower. Therefore, more energy is given to seeds with high sequence coverage of target $m$ (\ie \texttt{a-c-e-i-l}). In this way, LeoFuzz has more chances to explore the left branch of Fig. \ref{icfg}, thus more likely to reach target $m$. 2) an adaptive exploration and exploitation coordination approach, which is based on two queues storing directed seeds and coverage seeds respectively. LeoFuzz coordinates exploration and exploitation stage flexibly according to the ratio of seeds in two queues.

LeoFuzz also uses a concolic executor to solve difficult constraints, such as magic number, which enables LeoFuzz to reach targets quickly. In addition, LeoFuzz combines call graph (CG) and control flow graph (CFG) to increase the length of each target sequence, thus further improving LeoFuzz's guidance on reaching targets.

\section{Approach}\label{approach}
\begin{figure}[t]
\centering
\includegraphics[width=\linewidth]{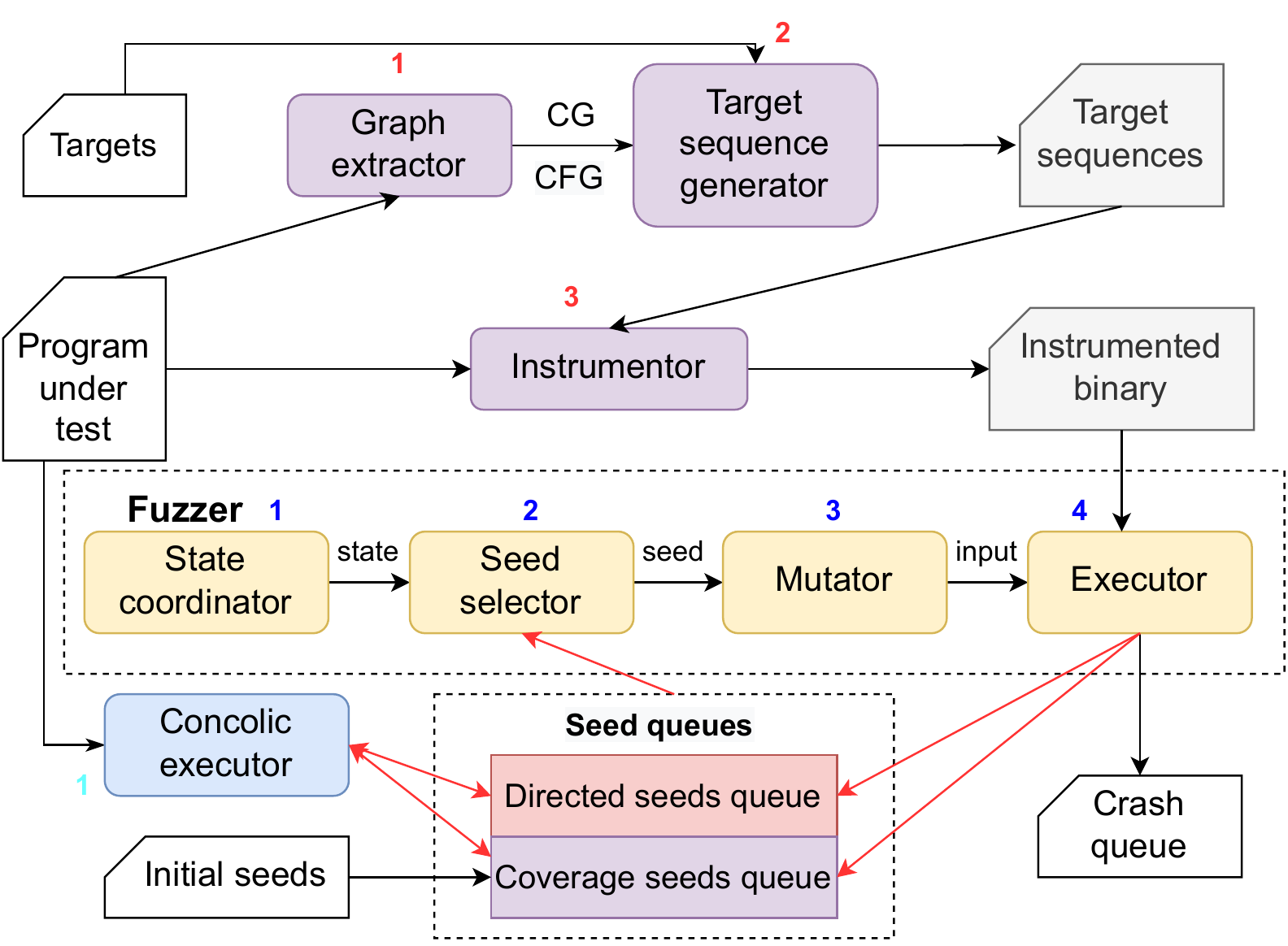}
\caption{LeoFuzz's architecture}
\label{LeoFuzz}
\end{figure}
The architecture of LeoFuzz is shown in Fig. \ref{LeoFuzz}, which includes two phases, \ie static analysis and dynamic analysis. In the static analysis phase, the graph extractor extracts a CG and a set of CFG from the program under test (PUT). Then the target sequence generator maps the statements in targets to basic blocks of the graphs and generates for each target a target sequence, which contains \emph{necessary} basic blocks along the paths to the target. Finally, the PUT is instrumented for collecting runtime information, such as code coverage and execution traces, and the instrumented binary is sent to the executor.

In the dynamic analysis phase, the fuzzer takes the initial seeds and the instrumented binary as inputs. First, the stage coordinator judges the stage of the fuzzer (exploration stage or exploitation stage). According to the stage, the seed selector then obtains a seed from the corresponding seeds queue, \ie coverage seeds queue (CQ for short) or directed seeds queue (DQ for short). After the mutator mutates the seed, the generated input is fed to the executor. The input is stored into the crash queue if it crashes the PUT, or into DQ if it increases the target sequence coverage, or into CQ if it increases code coverage, otherwise discarded. The fuzzer communicates with the concolic executor by sharing two seed queues. The concolic executor helps LeoFuzz focus on the paths going through the targets and explore more branches thus obtaining better code coverage. The concolic executor obtains seeds from two queues and stores its generated directed inputs or coverage inputs into the corresponding queue.

\section{Static analysis}\label{static-analysis}
\subsection{Generating Target Sequences}
Given a target, a target sequence is composed of a set of nodes, and each node is a \emph{necessary} basic block which exists on \emph{all} paths from the entry function (\eg \texttt{main}) to the target.
To make the fuzzer have better guidance, LeoFuzz combines CG and CFG of the PUT to enhance the target sequences, unlike Lolly and Berry, which only use CFG to generate target sequences.

  \begin{algorithm}[t]
\small
    \caption{Target Sequence Generation}
    \begin{algorithmic}[1]
      \REQUIRE Program $P$, Target $T$
      \ENSURE Target Sequence $TS$
        \STATE{TS = $\varnothing$;}
      \STATE{cg = getCG(P);}
      \STATE{domCg =  CG-to-Dom(cg);}
         \STATE{cfg = getCFG(P, T);}
        \STATE{domCfg = CFG-to-Dom(cfg);}
        \STATE{funName = getFunName(P, T);}
        \STATE{seq1 = getNesNodesByCFG(domCfg, T);}
        \STATE{seq2 = getNesNodesByCG(domCg, funName);}
        \STATE{TS = TS $\cup$ seq2 $\cup$ seq1;}

    \end{algorithmic}
      \label{alg:tsg}
  \end{algorithm}

As shown in Algorithm \ref{alg:tsg}, we generate the target sequence for a target based on Dominator Tree\footnote{A tree is called a dominating tree if each node in the tree \emph{dominates} only itself and its descendants. A node $d$ dominates a node $n$ if and only if each path that goes through $n$ must go through $d$ first. Each node dominates itself by definition.}~\cite{domtree}.  It takes a program $P$ and a target $T$ as inputs, and outputs the target sequence $TS$ which is initially empty. The algorithm first obtains CG from $P$ and converts it to the dominator tree (\ie $domCg$) (lines 2-3). Then we get CFG and convert it to the dominator tree $domCfg$ and get the target function name $funName$ (lines 4-6). Finally, we get the necessary nodes from $domCfg$ and $domCg$ respectively, and add them to $TS$ (lines 7-9).

\begin{figure}[t]
\centering
\includegraphics[height=9cm,width=9cm]{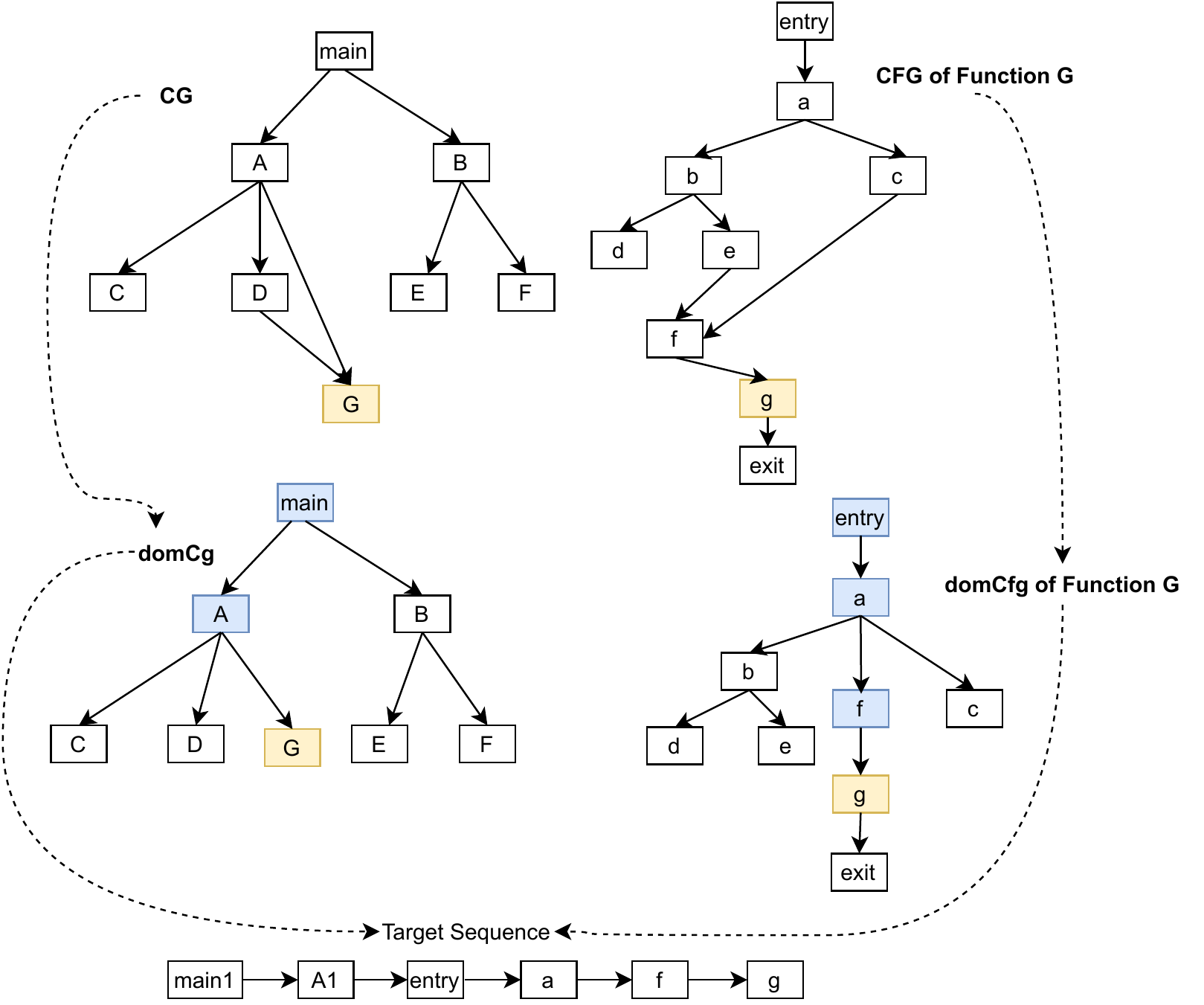}
\caption{Example: Construct target sequence based on CFG and CG}
\label{example1}
\end{figure}

We use an example to illustrate the algorithm, as shown in figure \ref{example1}. The figure shows the CG of a program $P$ and the CFG of function $G$ which contains a target $g$, as well as the dominating trees $domCg$ and $domCfg$. The blue nodes in the figure represent the \emph{necessary} nodes to reach the target $g$. Using Algorithm \ref{alg:tsg}, we can know that \texttt{main1-A1-entry-a-f-g} is the target sequence of the target $g$, where \texttt{main1} and \texttt{A1} represents the entry node of \texttt{main} and \texttt{A} function, respectively.

\subsection{Static Instrumentation}
Like Lolly, LeoFuzz instruments the basic blocks each of which contains at least a target statement, and uses a shared memory to sequentially record \emph{identifiers} of the blocks following the order in which they are executed (\ie execution trace).
As a result, the fuzzer can collect the code coverage information and execution traces related to targets during execution. These runtime information assists LeoFuzz in energy scheduling and exploration-exploitation coordination.

\section{Dynamic analysis}\label{dynamic-analysis}
During the dynamic analysis stage, the fuzzer and the concolic executor independently executes the program under test while they help each other by sharing the coverage seed queue (CQ) and the directed seed queue (DQ).

\begin{table}[t]
\renewcommand\arraystretch{1.2}
    \centering
    \caption{Description of symbols in Algorithm \ref{alg:mtdgf} \& \ref{alg:cee}.}
    \label{tab-symbol}
    \begin{tabular}{l p{6.5cm}}
    \hline
        Symbol & Description \\ \hline
        $sof$ &  stage of Fuzzer (0 for exploration, 1 for exploitation) \\ \hline
        $dsc$ &  number of seeds  in directed seed queue \\ \hline
        $csc$ &  number of seeds  in coverage seed queue \\ \hline
        $ndc$ &  number of consecutive executions in \emph{current} exploitation stage during each of which no directed seed is generated  \\ \hline
        $cdsc$ &  number of directed seeds generated in \emph{current} exploitation stage \\ \hline
        $rate$ & control coefficient to switch Fuzzer from exploration to exploitation stage\\ \hline
        $epoch$ &  frequency that Fuzzer switches to exploitation stage \\ \hline
        $lndc,slndc$ &  values of $ndc$ in the last two epochs \\ \hline
    \end{tabular}
\end{table}

\begin{algorithm}[t]
\small
    \caption{Multiple Targets Directed Grey-box Fuzzing}
    \begin{algorithmic}[1]
      \REQUIRE Instrumented Binary $P$, Target Sequence Set $TSS$, Coverage seed queue $CQ$, Directed Seed queue $DQ$,
      \ENSURE Crash seeds $CS$
    \STATE$sof = 0;  epoch = 0; dsc = 0;  csc = 0; ndc = 0; cdsc = 0;$
    \REPEAT
      \STATE $sof = stageCoord(sof)$;
%       \COMMENT{// using Alg.\ref{alg:cee}}
      \IF{$sof == 0$} 
%      \COMMENT{// in exploration stage}
        \STATE $s = getNextSeed(CQ)$;
      \ELSE 
%      \COMMENT{// in exploitation stage}
        \STATE $s = getNextSeed(DQ);$
      \ENDIF
      \STATE $p = assignEnergy(s);$
%       \COMMENT{// using MES strategy}
      \FOR{i from 1 to p}
        \STATE{$s' = mutate(s);$}
        \STATE{$execute(P,s');$}
        \IF{$s'$ crashes P}
        \STATE{add $s'$ to CS}
        \ELSE
        \STATE{$tseqCov = isIncreaseTSeqCov(s', TSS);$}
        \STATE{$codeCov = isIncreaseCodeCov(s');$}
        \IF{tseqCov} 
%        \COMMENT{// $s'$ increases target sequence coverage}
        \STATE add $s'$ to DQ;
        \STATE $dsc\textit{++}; cdsc\textit{++}; ndc = 0;$ continue;
        \ENDIF
        \IF{codeCov}
%        \COMMENT{// $s'$ increases code coverage}
        \STATE add $s'$ to CQ;
        \STATE $csc\textit{++}; ndc\textit{++};$ continue;
        \ENDIF
        \STATE{$ndc\textit{++};$}
        \ENDIF
      \ENDFOR
    \UNTIL{timeout or abort}
    \end{algorithmic}
    \label{alg:mtdgf}
  \end{algorithm}

\subsection{Fuzzer}
The fuzzer in LeoFuzz works like other DGF tools though we enhance it with two novel techniques. We propose a novel approach to adaptively coordinate exploration and exploitation stages (CEE for short) based on two queues and a novel energy scheduling strategy that considers more relations between seeds and targets (MES for short).

The fuzzer's workflow is shown in Algorithm \ref{alg:mtdgf}. Its inputs are the instrumented binary $P$, coverage seed queue $CQ$, directed seed queue $DQ$ and target sequence set $TSS$, and the output is a set of crash seeds $CS$. 
Table \ref{tab-symbol} describes the meaning of each symbol/variable used in Algorithm \ref{alg:mtdgf} and \ref{alg:cee}.
% $sof$ indicts the stage of fuzzer (0 for exploration, 1 for exploitation). $dsc$ and $csc$ is the number of seeds storing in $DQ$ and $CQ$ respectively. $ndc$ represents the number of consecutive executions in \emph{current} exploitation stage during each of which no directed seed is generated. $cdsc$ means the number of directed seeds generated in \emph{current} exploitation stage.

After initialization, the fuzzer decides its stage using CEE Algorithm \ref{alg:cee} and accordingly selects a seed $s$ from CQ or DQ respectively (lines 3-8).  It then assigns energy $p$ to the seed using MES strategy. In the energy loop, the fuzzer generates a new input $s'$ via mutation, and executes $P$ with $s'$. If $s'$ causes the program crash, increases code coverage or target sequence coverage, the fuzzer stores it in CS, CQ or DQ respectively (lines 13-25). Note that if $s'$ increase both code coverage and target sequence coverage, we store it in DQ since directed seeds are usually less than coverage seeds.

  \renewcommand{\algorithmicensure}{\textbf{return}}
  \renewcommand{\algorithmicrequire}{\textbf{function}}
  \begin{algorithm}[t]
\small
    \caption{Coordinating Exploration and Exploitation Stage}
    \begin{algorithmic}[1]
    \REQUIRE stageCoord(sof)
    \IF{$sof == 0$}
      \IF{$csc / (csc + dsc) > rate$}
      \STATE{$sof = 1;$} \COMMENT{// switch to exploitation}
      \STATE{$ndc = 0;$}
      \STATE{$cdsc = 0;$}
       \STATE{$epoch\textit{++};$}
      \ENDIF
      \ELSE
      \STATE{$th =1/2* (slndc+lndc) * \sqrt{epoch};$}
      \IF{$ndc >= th $}
      \STATE{$slndc = lndc;$} \COMMENT{// update ndc of the second last epoch}
      \STATE{$lndc = ndc;$} \COMMENT{// update ndc of the last epoch}
      \STATE{$updateRate(rate,cdsc,epoch);$}
      \STATE{$sof = 0;$} \COMMENT{// switch to exploration}
      \ENDIF
      \ENDIF
      \ENSURE $sof$
%      \ENSURE{$sof$}
    \end{algorithmic}
    \label{alg:cee}
  \end{algorithm}

\subsubsection{Exploration-exploitation Coordination}

CEE mechanism is shown in Algorithm \ref{alg:cee}. The fuzzer starts in the exploration stage initially and will switch to exploitation stage when the ratio of coverage seeds ($csc$) in total seeds is greater than a \emph{dynamic} threshold $rate$, which means that fuzzer has adequate code coverage information, so we set $sof$ as exploitation stage, and updates the related data, \eg $ndc, cdsc, epoch$ (lines 1-7). Note that we set the initial value of $dsc$ as 10 at the first exploration stage to prevent the fuzzer from immediately switching to exploitation.

In exploitation stage, we record related runtime information, \eg  $cdsc$ and $ndc$, as shown in Algorithm \ref{alg:mtdgf}. The fuzzer will switch to exploration stage when $ndc$ exceeds a threshold $th$, which indicates that the fuzzer's exploitation ability is weak now. Therefore we set $sof$ as exploration stage in order to explore more code paths, and \emph{update} the coefficient $rate$ (lines 8-16). The threshold $th$ is calculated by the values of $ndc$ in last two epochs and $epoch$ (line 9).  Because the probability of finding a new directed seed decreases gradually during fuzzing,  we increase $th$ at each epoch to keep the fuzzer in exploitation stage longer.

The $rate$ is used to decide when the fuzzer switches from the exploration stage to the exploitation stage, so we update its value by using run-time information in the current exploitation stage (\eg $cdsc$) only before the fuzzer will leave for the exploration stage. Obviously, the greater the $rate$ is, the longer the fuzzer stays in exploration, and vise versa.
To balance the fuzzer's code coverage and directed exploitation, we use function $updateRate$ to adjust $rate$ according to $epoch$ and $cdsc$, as follows: %make stage coordination more flexible
\begin{equation}
 rate^{*} = rate - \gamma(tanh(\frac{cdsc}{\sqrt{t}}*\sqrt{epoch}) - \delta)
 \label{eq1}
\end{equation}
where $rate^{*}$ indicts the new value of $rate$ for use in next exploration stage, $tanh()$ is a hyperbolic function, $t$ represents the time duration (seconds) of current exploitation stage. (Note: the upper and lower bound of $rate$ is set to 1 and 0, respectively.)

Specifically,  the larger $\frac{cdsc}{\sqrt{t}}$ is, which means that more directed seeds are produced in exploitation stage and that current code coverage is helpful in reaching targets,  the less time the fuzzer would use to reach targets and should switch to exploitation stage as soon as possible. As such, we reduce $rate$ to make the fuzzer enter exploitation stage fast.
On the contrary,  the smaller $\frac{cdsc}{\sqrt{t}}$ is, which indicates that current code coverage does not help reach targets, the more exploration time is needed to get more coverage information, therefore we increase $rate$ to keep the fuzzer exploring longer.
In addition, the probability of finding a new directed seed decreases gradually during fuzzing, resulting in a smaller $\frac{cdsc}{\sqrt{t}}$ and a greater $rate$ over time. Therefore, we use the parameter $epoch$ to offset this effect so that $rate$ changes reasonably.

\subsubsection{Seed Energy Scheduling for Multiple Targets}
To balance the energy scheduling for multiple targets, we propose a novel energy scheduling strategy (MES for short) that considers more relations between seeds and targets.

Specifically, for a seed $s$ and multiple targets, LeoFuzz generates all target sequences (\eg $N$ in total) at the static analysis phase (Section \ref{static-analysis}) and obtains the seed's execution trace during fuzzing. We consider three relations between the seed and target sequences as follows.

\begin{itemize}
\item The priority of a target sequence $TS_i$, which indicts $TS_i$'s similarity with other sequences and is computed at static analysis phase as follows:
\begin{equation}
%\footnotesize {
%priority_i = \sum_{j=1}^{N}{\left\{ \Bigl(\frac{LCS(TS_i, TS_j)}{Max(TS_i, TS_j)} \ge 0.4 \ ?~1 : 0\Bigr)   \land (j \ne i )\right\}}
% \begin{split}
priority_i = \sum_{j=1}^{N}{\Bigl(\frac{LCS(TS_i, TS_j)}{Max(TS_i, TS_j)} \ge \epsilon \ ?1 : 0\Bigr),  j \ne i } 
% \end{split}
%}
\label{eq2}
\end{equation}
where $LCS()$ indicts the length of the longest common subsequence of two sequences, $Max()$ returns the maximum length of two sequences. The higher the priority, the greater the chance that the fuzzer can reach \emph{multiple} targets by mutating $s$, so $s$ should be assigned more energy.

\item The global maximum coverage $gMaxCov_i$ of a target sequence $TS_i$. It refers to the maximum coverage of \emph{any} execution trace in the past over the target sequence, which approximates $TS_i$'s difficulty to cover and is updated during dynamic analysis. (In fact, the $gMaxCov$ used for the next seed energy scheduling is obtained by calculating the maximum value of the $seqCov$ of the current seed and the current $gMaxCov$.) The less $gMaxCov_i$ is, $TS_i$ is more difficult to be covered, the target corresponding to $TS_i$ is more difficult to reach, and thus the seed should be assigned more energy.

\item The seed's sequence coverage ($seqCov$) over a target sequence ($TS_i$). It measures the similarity between the seed's execution trace (ET) and $TS_i$, and is calculated during dynamic analysis as follows:
\begin{equation}
seqCov_i=\frac{LCS(ET,TS_i)}{lengh(TS_i)}
\end{equation}
where LCS() gets the length of the longest common subsequence between $s$' execution trace and $TS_i$. The greater the $seqCov_i$, the more likely that the fuzzer will cover $TS_i$ by mutating $s$, so $s$ should be assigned more energy.
\end{itemize}

For a seed and multiple targets, MES selects the target sequence\footnote{If there are multiple ones, the first is used.} with the maximum value of $seqCov$ as the seed's outstanding target sequence ($OTS$ for short), and considers $OTS$' $priority$, $gMaxCov$ and the seed's $seqCov$ over $OTS$ when scheduling energy for the seed.
In this way, LeoFuzz can provide a fine-grained energy scheduling and thus effectively improve the ability of DGF to cover multiple targets.

Below we show how to calculate these values by using an example. Given three target sequences, $TS_1$: \texttt{a-b-c-d-f-g},  $TS_2$: \texttt{a-b-c-g-h}, $TS_3$: \texttt{a-g-i-k}, we first calculate the priority of each target sequence as follows. LCS($TS_1$,$TS_2$) and LCS($TS_1$,$TS_3$) is 3 and 1 respectively, Max($TS_1$,$TS_2$) and Max($TS_1$,$TS_3$) are 6, so the priority of $TS_1$ is 1 according to equation (2). Similarly the priority of $TS_2$ and $TS_3$ is 1 and 0 respectively. Suppose that seed $s$' execution trace (ET) is \texttt{a-b-c-g-k-m-d}, we can get LCS($ET$,$TS_1$) is 3, and $s$' sequence coverage over $TS_1$  is $3/6 = 0.5$ according to equation (2). Similarly that of $s$ over $TS_2$, $TS_3$ is 0.8 and 0.25 respectively. Therefore, $s$' sequence coverage is 0.8, and  $s$' $OTS$ is $TS_2$.  Assuming that three seeds in total were executed in the past and their coverage with $TS_2$ is 0.3, 0.5 and 0.4 respectively, then the global maximum coverage of $TS_2$ is 0.5.

It is arduous to judge which target is difficult to reach when the fuzzer has insufficient code coverage, especially in initial executions. Therefore, we won't consider global maximum coverage of the target sequence in energy scheduling until the fuzzer has sufficient code coverage. For example, when the number of target sequences whose global maximum coverage is greater than or equal to a threshold ($\beta$) exceeds half of total sequences, the fuzzer likely has covered shallow (or easy) targets, it is reasonable to consider those deep (or difficult) targets. Therefore, those seeds corresponding to these targets will be allocated less energy and other targets have more opportunities to be explored.

We use a comprehensive factor (CF) to represent the above relations:
\begin{equation}
    CF=
\begin{cases}
\frac{1}{2}(seqCov + priority/N), & \\
\text{~~~~if \   $\sum_{j=1}^{N}{(gMaxCov_j \ge \beta~?~1 : 0}) < \frac{1}{2}N$}\\
\frac{1}{3}(seqCov + priority/N + (1 - gMaxCov)),& \\
\text{~~~~if \   $\sum_{j=1}^{N}{(gMaxCov_j \ge \beta~?~1 : 0}) \ge \frac{1}{2}N$}
\end{cases}
\end{equation}
where N is the number of total sequences.

AFLGo uses an energy scheduling scheme based on simulated annealing in the greybox fuzzing. Different from the traditional random walk algorithm which may be trapped in a local optimum, the simulated annealing algorithm accepts a solution which is worse than the current one with a certain probability, so it can jump out of the local optimum and reach the global optimum. This probability gradually decreases as the control parameter \emph{temperature} decreases.

Like AFLGo, LeoFuzz also applies simulated annealing to our energy scheduling for a global optimum and uses the same coefficient values as AFLGo in the following equations. For multiple targets directed fuzzing, an optimal solution is a test case that can achieve the maximum  CF.  In our method, the temperature $T$ with an initial value $T_0$= 1 is exponentially cooled.
\begin{equation}
    T = T_0 \times \alpha ^ k
\end{equation}
where, $\alpha$ is a constant which meets $ 0.8 \leq \alpha \leq 0.99$, and $k$ is the temperature cycle. The threshold of $T_k$ is set to 0.05. The fuzzer will not accept worse solutions when the temperature is lower than $T_k$. Specifically, if $T_k > 0.05$, LeoFuzz randomly mutates the existing seeds to generate many new inputs. Otherwise, it generates more new inputs from seeds with higher CF. In this case, the simulated annealing process is similar to the traditional gradient descent algorithm.

Since a common limitation of fuzzing is the time budget, we use time $t$ to adjust the temperature cycle $k$:
\begin{equation}
    \frac{k}{k_x} = \frac{t}{t_x}
\end{equation}

where $k_x$ and $t_x$ are the temperature cycle and the time respectively when temperature drops to $T_k$. So we use $k$ to establish the relationship between time $t$ and temperature $T$:
\begin{equation}
    T_k = 0.05 = \alpha^{k_x}
\end{equation}

\begin{equation}
    T = \alpha^k = \alpha^{t/t_x \times \log(0.05)/\log{(\alpha)}} = 20^{-{t/t_x}}
\end{equation}

Given a seed $s$, multiple targets, and their comprehensive factor (CF), we define the capability of $s$ to cover the given the multiple targets as:
\begin{equation}
    Cap = CF\times(1-T)+0.5\times T
\end{equation}

At the beginning of fuzzing, the initial value of temperature $T$ is 1, which means that $Cap$ is independent of $CF$. As time goes on, the temperature $T$ gradually decreases and $CF$ becomes increasingly important.

To combine our MES strategy with the existing seed energy schedule algorithm of a fuzzer (\eg  AFL), LeoFuzz integrates the capability of covering multiple targets ($Cap$) into the energy calculation formula. LeoFuzz calculates the integrated energy for a seed as:
\begin{equation}
    Menergy = energy\times2.0^{(Cap - 0.2) \times 10}
\end{equation}
where $energy$ is the original energy given by AFL and $Menergy$ is the energy given by LeoFuzz which integrates MES strategy in the original fuzzer.

\subsection{Concolic Executor}
The fuzzer leverages a random mutation to generate test inputs without considering the context of the PUT and thus it has difficulty to reach deep targets and find deep errors along complex paths ~\cite{vuzzer,angora,aflfast,fairfuzz}. Therefore, we combine the fuzzer with the concolic executor to solve this problem. The concolic executor continuously obtains seeds from two seed queues, executes them,  and generates new inputs, which are then put into CQ or DQ if they bring new code coverage or new target sequence coverage. To generate directed seeds as quickly as possible so that LeoFuzz can reach the targets faster, the concolic executor prefers to acquire seeds from DQ first and then seeds from CQ if there are no available directed seeds.

\section{Implementation}\label{Implement}
\textbf{Static analysis:} We wrote an LLVM pass which builds call graph (CG) for the program under test and control flow graph (CFG) for each of its functions. The dominator tree is constructed for each CFG and CG by NetworkX, and the necessary nodes are calculated to obtain the target sequence. We modified the AFL's instrumentation pass which writes the IDs of basic blocks in a target sequence into the shared memory, in order to record the target sequence's execution trace.

\textbf{Dynamic analysis:} We implemented our exploration-exploitation coordination strategy and energy scheduling strategy in AFL. In addition, we modified the concolic executor in QSYM to leverage guidance from the directed seeds, and hence LeoFuzz can reach the targets faster.
%help LeoFuzz reach the target faster by generating seeds covering the target path with complex constraints by using the concolic executor and the directed seeds generated by the fuzzer. 

\begin{table}
\renewcommand\arraystretch{1.2}
    \centering
    \caption{Subject programs for crash reproduction. IFT: Input File Type}
    \label{tab:put}
%    \begin{tabular}{c|c|c|c|c}
    \begin{tabular}{c|c|c|c|C{2.5cm}}  
    \hline
        Package & Program & IFT & \#CVE & Arguments \\  \hline
       & cxxfilt & ELF &5 &  \\ \cline{2-5}
       binutils\hfill\cite{binutils} & objdump & ELF & 4 & --dwarf-check -C -g -f -dwarf -x @@  \\ \cline{2-5}
        & readelf & ELF &2 & -zR3 @@  \\ \cline{2-5}
        libtiff\hfill \cite{libtiff}& tiff2pdf & TIFF &2 & @@  \\\hline
        libredwg\hfill \cite{libredwg}& dwg2dxf & DWG &10 & @@ -o /dev/null  \\\hline
        zziplib\hfill \cite{zziplib}& unzzipcat-mem & ZIP  &6 & @@  \\\hline
        mjs\hfill \cite{mjs}& mjs & JS &2 &   \\\hline
    \end{tabular}
\end{table}
\section{Evaluation}\label{evaluation}
In this section, we first evaluate LeoFuzz's effectiveness and efficiency in terms of crash reproduction, true positive verification and vulnerability exposure in real-world programs, and compare the performance difference of running a LeoFuzz instance given multiple targets vs. running in parallel multiple LeoFuzz instances each of which aims to reach a target. Then we evaluate the contributions of four main design decisions in LeoFuzz, \ie target sequence enhancement, exploration-exploitation coordination, fine-grained energy scheduling strategy and concolic execution, respectively.

\subsection{Experiment Setup}
We conducted all experiments on a virtual machine with an Intel(R) Xeon(R) Gold 6126 CPU, 128GB RAM and Ubuntu 18.04 (64 bit) as operating system. To evaluate the effectiveness and efficiency of LeoFuzz, we compare it with six state-of-the-art fuzzers, AFLGo, Lolly, Berry, QSYM, Beacon and WindRanger. We evaluated them with the same programs under test, initial input corpus, target locations and time budget (12 hours). Since Beacon cannot support multiple targets, we ran it and compared it with LeoFuzz in experiments with single target. Note that Hawkeye \cite{hawkeye}, RDFuzz \cite{RDFuzz} and CAFL \cite{cafl} are not publicly available, so LeoFuzz doesn't compare with them. LeoFuzz was not compared to ParmeSan \cite{ParmeSan} because we could not replicate its experiments successfully even after we asked its authors for help.

We leverage seven real-world programs shown in Table \ref{tab:put} for all experiments except clearly stated,  because these programs are widely used, extensively evaluated by fuzzing tools in both academia~\cite{QSYM,hawkeye,libfuzzer} and industry~\cite{ossfuzz}, and found vulnerable due to multiple bugs.
For these programs, we collected 31 vulnerabilities and corresponding arguments from CVE database or their official sites and took each vulnerability's crash site as a target in experiments, \ie A unique target location corresponds to a unique vulnerability.
Furthermore, we repeated all experiments 10 times and used the average values.

In the experiments, we aim to answer the following questions:
\begin{enumerate}[itemindent=0.5em,label=\textbf{RQ\arabic*}]
\item How does LeoFuzz perform when it is given a single target each time?
\item Is LeoFuzz effective and efficient in crash reproduction? 
\item How does running LeoFuzz with multiple targets compared to running multiple LeoFuzz instances with one target per instance?
\item Is LeoFuzz efficient in terms of true positives verification?
\item Is LeoFuzz effective to discover vulnerabilities in real-world software?
\item How do four main design decisions contribute to LeoFuzz?
%techniques, \ie target sequence enhancement, exploration-exploitation coordination, fine-grained energy scheduling strategy and concolic execution, 
\end{enumerate}

\begin{table*}%{sidewaystable*} %{table*}
\renewcommand\arraystretch{1.2}
%\footnotesize
    \centering
    \caption{Results of LeoFuzz and baseline tools when given a single target.
    UAF=use-after-free, IO=integer overflow, BOF=buffer overflow, SOF=stack overflow, NP=null point exception, ML=memory leak, AE=arithmetic error, OR=out-of-bounds Read, IR=invalid memory read}
    \label{tab-crashrep}
    \begin{tabular}{p{4em}|l|l|l|r|l|r|l|r|l|r|l|r|l}
    \hline
         \multirow{2}{*}{Program}&\multirow{2}{*}{CVE-ID}  & \multirow{2}{*}{Type}  & \multicolumn{2}{c|}{QSYM} & \multicolumn{2}{c|}{AFLGo} & \multicolumn{2}{c|}{Lolly} & \multicolumn{2}{c|}{Berry} & \multicolumn{2}{c|}{Beacon} &  LeoFuzz$^s$  \\
         \cline{4-14}
         &   &   & TTE & Factor & TTE  &Factor & TTE & Factor & TTE & Factor & TTE & Factor &TTE  \\ \hline
         & 2016-4487 & UAF & 2m54s & 2.32 & 2m57s & 2.36 &3m12s&2.56&2m19s&1.86 & 2m50s& 2.27 & \textbf{1m15s} \\
         & 2016-4489 & IO & 3m53s & 2.04 & 2m12s & 1.16 &3m21s&1.76&2m27s&1.29& 2m51s &1.50 &\textbf{1m54s} \\
        cxxfilt & 2016-4490 & IO  & 50s& 1.56 & 50s & 1.56 &49s&1.53&42s&1.31 & 34s &1.06 &\textbf{32s} \\
         & 2016-4491 & BOF  &  1h23m & 1.69 & 1h46m & 2.16 &1h43m&2.08&1h19m&1.59& 1h26m & 1.73 & \textbf{49m36s} \\
         & 2016-4492 & BOF  &  2m15s & 1.07 & 4m41s & 2.23 &5m17s&2.52&4m52s&2.32&3m55s&1.87 &\textbf{2m6s} \\ \hline
        \multirow{4}{*} {objdump}& 2018-17985 & SOF  & --- & 3.43 &  --- & 3.43 &10h17m&2.95&7h41m&2.21&4h26m&1.27 &\textbf{3h29m} \\
         & 2018-20671 & BOF  & --- & 1.89 &  9h44m & 1.54 &---&1.89&7h27m&1.18 & --- &1.89&\textbf{6h18m} \\
        & 2018-9138 & SOF & --- & 4.11 & --- & 4.11 &---&4.11&---&4.11&2h59m&1.01 &\textbf{2h55m} \\
         & 2019-9070 & SOF & --- & 5.14 & 6h2m & 2.60 &6h53m&2.97&4h47m&2.06&2h39m&1.14 &\textbf{2h19m} \\ \hline
         \multirow{2}{*} {readelf} & 2017-7209 & NP &  6h0m & 4.53 & 2h14m & 2.53 &3h4m&3.50&1h56m&2.21&---&13.64 &\textbf{52m48s} \\
         & 2019-14444 & IO & 9h37m & 8.01 & 4h22m & 3.64 &3h51m&3.21&3h12m&2.67&---&10.00 &\textbf{1h12m} \\ \hline
       \multirow{2}{*} {tiff2pdf} & 2018-15209 & BOF & ---  & 11.80 & --- & 11.80 &---&11.80&4h43m&4.64&---&11.80 &\textbf{1h1m} \\
        & 2018-16335 & BOF  & --- & 7.06 & 4h14m & 2.49 &5h23m&3.17&4h13m&2.48 &---&7.06 &\textbf{1h42m} \\ \hline
        & 2019-9770 & BOF  & 8h7m  &  1.90 &  8h24m &  1.97 & 9h11m & 2.15 & 8h19m & 1.95 &---&2.81 & \textbf{4h16m} \\
        & 2019-9771 & NP  & 3h38m  &  17.37 &  31m41s &  2.53 & 35m11s & 2.81 & 27m49s & 2.22 &28m18s&2.25 & \textbf{12m33s} \\
        & 2019-9772 & NP & --- & 19.73 & 3h26m & 5.64 & 2h57m & 4.85 & 2h19m & 3.81 &---&19.73 &  \textbf{36m29s} \\
        & 2019-9773 & BOF & 7h57m & 1.82 & 8h35m & 1.97 & 9h17m & 2.13 & 8h13m & 2.04 &---&2.75 & \textbf{4h22m} \\
        & 2019-9774 & OR & 32m17s & 6.07 & 10m10s & 1.91 & 7m26s & 1.40 & 7m2s & 1.32 &22m37s&4.26 & \textbf{5m19s} \\
        dwg2dxf & 2019-9775 & OR & 2h2m & 3.25 & 6h5m & 9.72 & 5h1m & 8.02 & 3h32m & 5.65 &5h45m&9.18 & \textbf{37m35s} \\
        & 2019-9776 & NP & --- & 15.47 & 1h53m & 2.43 & 2h12m & 2.84 & 1h43m & 2.21 &49m21s& 1.06 & \textbf{46m33s} \\
        & 2019-9777 & BOF & 41s & 1.46 & 49s & 1.75 & 47s & 1.68 & 33s & 1.18  &31s &1.11 &\textbf{28s} \\
        & 2019-9778 & BOF & 33m32s & 1.22 & 42m57s & 1.57 & 47m31s & 1.73 & 39m32s & 1.44 & 31m9s &1.14 & \textbf{27m22s} \\
        & 2019-9779 & NP & 7h40m & 12.33& 59m47s & 1.60 & 53m24s & 1.41 & 43m44s & 1.17 &41m27s&1.10 &  \textbf{37m18s} \\ \hline
        & 2017-5974 & BOF & --- & 4.31 & 7h36m & 2.73 & 8h36m & 3.09 & 7h42m & 2.77 &---&4.31 &   \textbf{2h47m} \\
        & 2017-5975 & BOF & --- & 18.32 & 1h43m & 2.62 & 1h16m & 1.93 & 1h3m & 1.60 &---&18.32 &   \textbf{39m16s} \\
        unzzipcat & 2017-5976 & BOF & 3h57m & 11.67 & 38m50s & 1.91 & 41m22s & 2.04 & 26m49s & 1.32 &7h58m &23.55 &  \textbf{20m18s} \\
        -mem & 2017-5977 & IR & 3h45m & 10.09 & 33m26s & 1.34 & 37m43s & 1.39 & 31m4s & 1.14 &7h24m &16.26 & \textbf{27m18s} \\
        & 2017-5978 & OR & 2h57m & 6.44 & 1h23m  & 3.02 & 1h8m & 2.47 & 57m54s & 2.11 &29m28s&1.07 &  \textbf{27m31s} \\
        & 2017-5980 & NP & 8m17s & 2.49 & 6m17s  & 1.89 & 6m9s & 1.85 & 6m11s & 1.86 & 5m27s&1.64 &  \textbf{3m19s} \\ \hline
        \multirow{2}{*} {mjs} & issues-59 & AE &  3h50m & 15.33 & 34m1s & 2.24 &30m12s&1.99&31m54s&2.10&21m13s&1.39 &\textbf{15m12s} \\
         & issues-136 & SOF &  2h38m  & 15.05 & 10m29s & 1.00 &12m14s&1.18&11m3s&1.06&11m22s&1.09 &\textbf{10m27s} \\\hline
    \end{tabular}
\end{table*}%{sidewaystable*}{sidewaystable*} %{table*}
\subsection{Crash reproduction}
Software programs may crash due to potential bugs or vulnerabilities. A crash report usually contains memory dumps or call stacks of the program. Based on it, developers need to generate test cases that trigger the crash, \ie reproduce the crash. Directed fuzzing technique is demonstrated effective on crash reproduction \cite{AFLGo,Lolly,Berry}.

Using 31 vulnerabilities from seven real-world programs shown in Table \ref{tab:put}, we evaluate  LeoFuzz's ability on crash reproduction. We conduct experiments using two settings: 1) we run LeoFuzz with a single target (\ie RQ1), in this case, we name it LeoFuzz$^s$ for convenience, and 2) we run LeoFuzz with multiple targets (RQ2), comparing to baseline tools, respectively. Then we compare the difference between two settings of LeoFuzz (\ie RQ3).
\subsubsection{RQ1: Performance of LeoFuzz when given a single target}

The experimental results for RQ1 are shown in Table \ref{tab-crashrep}. The first column is the program under test. The second and third column is the vulnerability's identification and type respectively. Time-to-Exposure (TTE) measures the length of the fuzzing campaign until the first test input is generated that exposes a given vulnerability. Factor measures the performance gain as the mean TTE of each baseline tool divided by that of LeoFuzz$^s$. Values of factor greater than one indict that LeoFuzz$^s$ performs better than the corresponding baseline tool. Note that when a vulnerability is not found in the time budget (\ie 12 hours),  we use the time budget to calculate the factor. The fourth column to the thirteenth column are the mean of TTE and factor of QSYM, AFLGo, Lolly, Berry and Beacon in 10 experiments, respectively. The last column is the mean of TTE of LeoFuzz$^s$.

As shown in Table \ref{tab-crashrep}, LeoFuzz$^s$ can reproduce each crash while Beacon, QSYM, AFLGo, Lolly and Berry fails on 10, 8, 3, 2 vulnerabilities and 1 vulnerability, respectively. Moreover,  LeoFuzzer$^s$ is faster than all baseline tools, \ie 7.12$\times$ than QSYM, 2.92$\times$ than AFLGo, 2.83$\times$ than Lolly, 2.21$\times$ than Berry and 5.46$\times$ than Beacon, respectively.

\subsubsection{RQ2: Effectiveness and efficiency of LeoFuzz when given multiple targets}
\begin{table}[t]
\renewcommand\arraystretch{1.2}
    \centering
    \caption{Results of LeoFuzz and baseline tools when given multiple targets}
    \label{tab-multargets}
    \begin{tabular}{p{0.08\columnwidth}|p{0.12\columnwidth}|c|c|c|l|r|r}
    \hline
        Prog. & Tool & Tgt. & Rep. & Add. & TTE  & Factor & $A_{12}$ \\ \hline
         & {\scriptsize QSYM} & 5 & 5 & 1 & 1h23m & 1.79 & 0.59\\
        &{\scriptsize AFLGo} & 5 & 5 & 1 & 1h54m  & 2.46 & 0.54\\
     {\scriptsize cxxfilt}  & {\scriptsize Lolly} & 5 & 5 & 1 & 1h37m & 2.09 & 0.62 \\
        & {\scriptsize Berry} & 5 & 5 & 2 & 1h21m & 1.75 & 0.55\\
        & {\scriptsize WindRanger} & 5 & 5 & 2 & \textbf{39m50s} & 0.86 & 0.41 \\
        & {\scriptsize LeoFuzz} & 5 & 5 & \textbf{3} & 46m22s & ---& --- \\ \hline
         & {\scriptsize QSYM} & 4 & 0 & 1 &  ---  & 2.90& 1.00\\
        & {\scriptsize AFLGo} & 4 & 3 & 1 &  --- & 2.90& 1.00 \\
     {\scriptsize  objdump} & {\scriptsize Lolly} & 4 & 3 & 2 &  --- & 2.90& 1.00\\
        & {\scriptsize Berry} & 4 & 3 & 3 &  --- & 2.90& 1.00\\
        & {\scriptsize WindRanger} & 4 & 1 & 0 & --- & 2.90& 1.00 \\
         & {\scriptsize LeoFuzz} & 4 & \textbf{4} & \textbf{5} & \textbf{4h28m} & ---& --- \\ \hline
         & {\scriptsize QSYM} & 2 & 2 & 1 & 9h37m & 10.27 & 0.83\\
        & {\scriptsize AFLGo} & 2 & 2 & 1 & 5h21m & 5.71 & 0.95\\
      {\scriptsize readelf} & {\scriptsize Lolly} & 2 & 2 & 1 & 5h49m & 6.21 & 0.94\\
        & {\scriptsize Berry} & 2 & 2 & 1 & 3h17m & 3.51 & 0.87\\
        & {\scriptsize WindRanger} & 2 & 2 & 0 & 4h11m & 4.47 & 0.96 \\
         & {\scriptsize LeoFuzz} & 2 & 2 & 1 & \textbf{56m11s} & --- & ---\\ \hline
         & {\scriptsize QSYM} & 2 & 0 & 0 &  --- & 9.35 & 1.00\\
        & {\scriptsize AFLGo} & 2 & 0 & 1 &  --- & 9.35 & 1.00 \\
     {\scriptsize  tiff2pdf} & {\scriptsize Lolly} & 2 & 1 & \textbf{1} &  --- & 9.35 & 1.00 \\
        & {\scriptsize Berry} & 2 & \textbf{2} & 0 &  4h23m & 4.06 & 0.81\\
        & {\scriptsize WindRanger} & 2 & 1 & 0 & --- & 9.35 & 1.00 \\
        & {\scriptsize LeoFuzz} & 2 & \textbf{2} & \textbf{1} & \textbf{1h17m} & --- & --- \\ \hline

         & {\scriptsize QSYM} & 10 & 8 & \textbf{6} &  --- & 3.12 & 1.00\\
        & {\scriptsize AFLGo} & 10 & 7 & 3 &  --- & 3.12 &1.00\\
     {\scriptsize  dwg2dxf} & {\scriptsize Lolly} & 10 & 7 & 3 &  --- & 3.12 & 1.00 \\
        & {\scriptsize Berry} & 10 & 7 & 4 &  --- & 3.12 & 1.00\\
        & {\scriptsize WindRanger} & 10 & 7 & \textbf{6} & --- & 3.12 & 1.00 \\
        & {\scriptsize LeoFuzz} & 10 & \textbf{10} & \textbf{6} & \textbf{3h51m} & --- & ---\\ \hline

         & {\scriptsize QSYM} & 6 & 4 & 1 &  --- & 6.99 & 1.00\\
        & {\scriptsize AFLGo} & 6 & 5 & \textbf{2} &  --- & 6.99 &1.00 \\
      {\scriptsize unzzipc} & {\scriptsize Lolly} & 6 & 4 & \textbf{2} &  --- & 6.99 & 1.00\\
      {\scriptsize atmem} & {\scriptsize Berry} & 6 & 5 & \textbf{2} &  --- & 6.99 & 1.00\\
       & {\scriptsize WindRanger} & 6 & 4 & \textbf{2} & --- & 6.99 & 1.00 \\
        & {\scriptsize LeoFuzz} & 6 & \textbf{6} & \textbf{2} & \textbf{1h43m} & --- & ---\\ \hline

         & {\scriptsize QSYM} & 2 & 2 & 1 &  3h50m & 17.10 & 1.00\\
        & {\scriptsize AFLGo} & 2 & 2 & 1 &  25m42s & 1.91 & 0.81\\
   {\scriptsize  mjs}   & {\scriptsize Lolly} & 2 & 2 & 2 &  21m52s & 1.62 & 0.75\\
        & {\scriptsize Berry} & 2 & 2 & \textbf{4} &  24m11s & 1.80 & 0.80 \\
        & {\scriptsize WindRanger} & 2 & 2 & 1 & 47m17s & 3.51 & 0.98 \\
         & {\scriptsize LeoFuzz} & 2 & 2 & \textbf{4} &  \textbf{13m27s} & --- & --- \\ \hline
    \end{tabular}
\end{table}

In this setting, we evaluate LeoFuzz's capability to deal with multiple targets. The baseline tools are also run with the same programs and targets as LeoFuzz.
Table \ref{tab-multargets} describes the experimental results. The first column is the program under test and the second column lists the fuzzing tools. The third column is the number of targets that are given to the tools, The fourth column is the number of bugs reproduced by each tool. The fifth column is the number of \emph{additional} bugs discovered by each tool.
The sixth column is the mean TTE spent by each tool to trigger \emph{all} specified bugs. The Vargha-Delaney statistic ($A_{12}$) is a standard measure for evaluating randomized algorithms \cite{a12}. Given a performance measure $M$ (\eg TTE ), the $A_{12}$ statistic measures the probability that running LeoFuzz yields higher $M$ values than running baseline tools, indicating the confidence that LeoFuzz performs better than the baseline tools. %seen in $m$ measures of LeoFuzz and $n$ measures of baseline tools

As shown in Table \ref{tab-multargets}, when given multiple targets, LeoFuzz can trigger all bugs within 4 hours, while QSYM, AFLGo, Lolly, Berry and WindRanger failed to trigger 10, 7, 7, 5 and 9 bugs within 12 hours, respectively. Moreover, LeoFuzz is able to discover more unique bugs in all programs than the baseline tools. In addition, LeoFuzzer is faster than all baseline tools, \ie 7.35$\times$ than QSYM, 4.63$\times$ than AFLGo, 4.61$\times$ than Lolly, 3.45$\times$ than Berry and 4.46$\times$ than WindRanger, respectively. Furthermore, LeoFuzz performs better with 92\%, 90\%, 90\%, 86\% and 91\% confidence on average than QSYM, AFLGo, Lolly, Berry and WindRanger, respectively.

\subsubsection{RQ3: One LeoFuzz with multiple targets versus multiple LeoFuzz$^s$ with a target per instance}

As shown in the previous sections, both LeoFuzz and LeoFuzz$^s$ are effective in crash reproduction. Then a question arises naturally; running LeoFuzz with multiple targets or running multiple LeoFuzz instances with one target per instance, which one is more efficient? We explore the question in this section and the experimental results are shown in Table \ref{tab:1vsn}. The third and fourth column is the sum and the longest of TTE spent by each LeoFuzz$^s$ instance to trigger a given vulnerability respectively. The fifth column measures the vulnerabilities triggered by LeoFuzz, and the sixth column is the TTE spent by LeoFuzz to trigger all vulnerabilities in each program.

As shown in Table \ref{tab:1vsn}, both one LeoFuzz instance and multiple LeoFuzz$^s$ instances can trigger all vulnerabilities, however, the time cost of LeoFuzz is less than the total time and even the longest time spent by each LeoFuzz$^s$ instance. For example, LeoFuzz spent 56m11s when triggering both CVE-2017-7209 and CVE-2019-14444 in \texttt{readelf} program, while LeoFuzz$^s$ took 1h12m when triggering CVE-2019-14444 only (see Table \ref{tab-crashrep}). The efficiency of LeoFuzz may benefit from the fact that real-world programs often have multiple bugs which are usually dependent or related, \eg caused by the same bad programming practices. In fact, we observed that the call stack in crash dump caused by CVE-2017-7209 has a large overlap with that caused by CVE-2019-14444, in other words, the paths to reach both vulnerabilities go through much same functions, which helps LeoFuzz trigger both of them fast. %and this relationship is intentional for the fuzzer to reach the target. The relationship between the multiple vulnerabilities of the program can further prove that it is meaningful for LeoFuzz to handle multiple targets.}

%We argue that the significant efficiency of LeoFuzz benefits from the fact that real-world programs often have multiple bugs.

\begin{table}[t]
\renewcommand\arraystretch{1.2}
    \centering
    \caption{One LeoFuzz versus multiple LeoFuzz$^s$ instances}
    \label{tab:1vsn}
    \begin{tabular}{l|c|p{0.12\columnwidth}|p{0.12\columnwidth}|c|p{0.1\columnwidth}}
    \hline
        Program & Tgt. &  LeoFuzz$^s$ & LeoFuzz$^s$ &  Rep. &  LeoFuzz\\
        &&Total TTE & Longest TTE & &TTE \\ \hline
        cxxfilt & 5 & 55m23s & 49m36s & 5 & 46m22s  \\
        objdump & 4 & 15h1m & 6h18m & 4 & 4h28m\\
        readelf & 2 & 2h4m & 1h12m & 2 & 56m11s\\
        tiff2pdf & 2 & 2h43m & 1h42m & 2 & 1h17m \\
        dwg2dxf & 10 & 13h1m & 4h22m & 10 & 3h51m  \\
        unzzipcat-mem& 6 & 4h25m  & 2h47m & 6 & 1h43m  \\
        mjs & 2 & 25m39s & 15m12s & 2 & 13m27s  \\\hline
    \end{tabular}
\end{table}

\subsection{RQ4: Efficiency on True Positives Verification}
\begin{table*} %sidewaystable
\renewcommand\arraystretch{1.2}
%\footnotesize
    \centering
    \caption{Results on True Positive Verification.}
    \label{tab-tpverify}
    \begin{tabular}{l|l|r|l|r|l|r|l|r|l}
    \hline
          \multirow{2}{*}{ CVE-ID}& \multicolumn{2}{c|}{QSYM} & \multicolumn{2}{c|}{AFLGo} & \multicolumn{2}{c|}{Lolly} & \multicolumn{2}{c|}{Berry} & LeoFuzz  \\
          \cline{2-10}
          & TTE & Factor & TTE  &Factor & TTE & Factor & TTE & Factor &TTE  \\ \hline
         2018-9132 & 7h19m & 6.55 & 7h55m & 7.09 & 6h11m & 5.54 & 5h48m & 5.19 & \textbf{1h7m} \\
         2018-9009 & 4h11m & 10.20 & 3h23m & 8.25 & 3h12m & 7.78 & 2h35m & 6.29 & \textbf{24m39s} \\
         2018-8807 & 25m43s & 8.60 & 9m30s & 3.06 & 6m11s & 2.04 & 4m18s & 1.42 & \textbf{3m2s} \\
         2018-7877 & 5h41m & 4.55 & 2h55m & 2.33 & 3h9m & 2.52 & 2h49m & 2.26 & \textbf{1h15m} \\
         2018-7876 & 2h16m & 3.00 & 3h54m & 5.14 & 2h13m & 2.92 & 2h42m & 3.56 & \textbf{45m23s} \\
         2018-7875 & 5h12m & 4.33 & 2h57m & 2.46 & 3h26m & 2.86 & 2h53m & 2.40 & \textbf{1h12m} \\
         2018-7873 & 5h59m & 6.71 & 3h2m & 3.40 & 2h31m & 2.88 & 1h47m & 2.00 & \textbf{53m21s} \\
         2018-7872 & 8h27m & 7.68 & 6h47m & 6.17 & 1h59m & 1.80 & 1h51m & 1.68 & \textbf{1h6m} \\
         2018-7870 & 5h38m & 6.30 & 3h12m & 3.48 & 2h34m & 2.79 & 1h54m & 2.07 & \textbf{55m12s} \\
         2018-7869 & 6h1m & 17.56 & 1h6m & 3.22 & 1h8m & 3.32 & 49m46s & 2.44 & \textbf{20m25s} \\
         2018-7868 & 8h1m & 10.41 & 6h13m & 8.07 & 1h42m & 2.21 & 1h18m & 1.69 & \textbf{46m12s} \\
         2018-7867 & 5h37m & 6.54 & 1h45m & 2.04 & 1h17m & 1.50 & 52m16s & 1.02 & \textbf{51m32s} \\
         2018-6359 & 41m52s & 5.25 & 18m4s & 2.36 & 13m43s & 1.79 & 13m38s & 1.79 & \textbf{7m39s} \\
         2018-6315 & 4m14s & 2.19 & 3m17s & 1.70 & 5m19s & 2.75 & 5m59s & 3.09 & \textbf{1m56s} \\
         2018-20591 & 4h6m & 13.92 & 3h13m & 10.92 & 2h35m & 8.77 & 2h15m & 7.64 & \textbf{17m41s} \\
         2018-20429 & 7h49m & 5.94 & 6h22m & 4.84 & 2h29m & 1.89 & 2h18m & 1.75 & \textbf{1h19m} \\
         2018-20427 & 7h6m & 5.20 & 8h2m & 5.88 & 6h49m & 4.99 & 6h0m & 4.39 & \textbf{1h22m} \\
         2018-11226 & 5h27m & 5.03 & 1h55m & 1.77 & 1h53m & 1.73 & 1h29m & 1.37 & \textbf{1h5m} \\
         2018-11225 & 9m45s & 5.48 & 5m20s & 3.24 & 4m12s & 2.42 & 3m3s & 1.76 & \textbf{1m44s} \\
         2018-11017 & 51m25s & 2.58 & 42m44s & 2.14 & 41m16s & 2.07 & 32m19s & 1.62 & \textbf{19m58s} \\\hline
    \end{tabular}
\end{table*}

Developers and testers usually apply static analysis tools to discover bugs or vulnerabilities in software before release. However, static analysis tools often have high false positive, and thus require a lot of manual efforts to verify their analysis results. Due to its directed execution feature, the DGF technique has been used for automatic verification of bugs~\cite{AFLGo, Lolly, Berry}. Moreover, Lolly and Berry outperformed over AFLGo due to their sequence coverage approach~\cite{Lolly, Berry}.

%\lhl{Reviewers may question us why we did not compare with Beacon and WindRanger in this RQ. If possible, please conduct experiments, otherwise, please explain the reasons.} 
We evaluated LeoFuzz’s ability on true positive verification and compared it with QSYM, AFLGo, Lolly and Berry. We use the same subject program, \ie Libming 0.4.8 ~\cite{libming}, as AFLGo, Lolly and Berry. In addition, we run the Clang Static Analyzer~\cite{clang} on the subject program and use its analysis results as targets, \ie the code locations of potential bugs. In the experiments, the analysis results of the Clang analyzer are not intentionally filtered and therefore may contain false positives and infeasible paths.
In order to evaluate the efficiency of LeoFuzz and four baseline tools, we guide them with the above targets to trigger CVE vulnerabilities of Libming and compare their time cost. The CVE vulnerabilities are listed in the first column of table \ref{tab-tpverify}.

Table \ref{tab-tpverify} presents the experimental results. The second to the ninth column is the mean of TTE and factor of QSYM, AFLGo, Lolly and Berry in ten runs, respectively. In particular, if a tool fails to trigger a vulnerability in a run within the time limit, its TTE is uniformly recorded as the time budget (\ie  12 hours). The last column is the mean TTE of LeoFuzz.

As shown in Table \ref{tab-tpverify}, five tools successfully generated inputs that can trigger the vulnerabilities, while LeoFuzz is 7.00$\times$ faster than QSYM, 4.45$\times$ than AFLGo, 3.33$\times$ than Lolly and 2.91$\times$ than Berry, respectively. Experimental results show that LeoFuzz is effective in true positives verification and more efficient than baseline tools.

\subsection{RQ5: Effectiveness on Vulnerabilities Exposure}
\begin{table*} %sidewaystable
\renewcommand\arraystretch{1.2}
%\footnotesize
    \centering
    \caption{Unreported Vulnerabilities Found by LeoFuzz.
    BOF=buffer overflow, SOF=stack overflow, NP=null point exception, MAF=memory allocation failure, UAF=use-after-free, ML=memory leak, BF=bad free, AE=assert error, DF=double free, SBOF=stack buffer overflow}
    \label{tab-newbugs}
    \begin{tabular}{l|p{4.5cm}|p{2.5em}|l} %{l|l|l|l|l} %
 \hline
        Program & Buggy Func. & Type &Reported as\\ \hline
        cxxfilt & demangle\_path & SOF & CVE-2021-3530 \\
         & demangle\_type  & SOF & ubuntu bug-1927070 \\ \hline
         & handleEditText & BOF & CVE-2021-42195 \\
         & traits\_parse & NP & CVE-2021-42196 \\
         & rfx\_alloc & ML & CVE-2021-42197 \\
         & swf\_GetBits & NP & CVE-2021-42198 \\
         & swf\_FontExtract\_DefineTextCallback & BOF & CVE-2021-42199 \\
         SWFTools & main  & NP & CVE-2021-42200 \\
         &  swf\_GetD64 & BOF & CVE-2021-42201 \\
         &  swf\_DeleteFilter & NP & CVE-2021-42202 \\
         & swf\_FontExtract\_DefineTextCallback & UAF & CVE-2021-42203 \\
         & swf\_GetBits & BOF & CVE-2021-42204 \\ \hline
         & bit\_calc\_CRC  & BOF & issues-484 \\
         & decode\_preR13 & NP & issues-485 \\
         & decode\_preR13\_section & NP & issues-486 \\
         & decode\_preR13\_section & UAF & issues-487 \\
         & decode\_preR13\_section\_hdr & BOF & issues-488 \\
libredwg & dwg\_add\_object  & BOF & issues-489 \\
         &  dwg\_add\_handleref & UAF & issues-490 \\
         &  dwg\_read\_file & BF & issues-491 \\
         & decode\_preR13\_entities & AE & issues-492 \\
         & dwg\_read\_file & DF & issues-493 \\
          & copy\_bytes & SBOF & issues-494 \\
         
         \hline
    \end{tabular}
\end{table*}
To evaluate LeoFuzz’s ability exposing bugs or vulnerabilities in real-world programs, we tested three widely used software with their latest versions, \ie cxxfilt 2.36 , SWFTools a9d5082~\cite{swftools} and 
libredwg 0.12.4.4608~\cite{libredwg}. Cxxfilt is a tool in Binutils, which decodes low-level names into user-level names to be human readable. SWFTools is a collection of utilities for working with Adobe Flash files. LibreDWG is a free C library to read and write DWG files.
The targets in this experiment come from the results of Clang static analyzer~\cite{clang} or the patches of the corresponding program under test,  and LeoFuzz aims to explore towards the potentially buggy code. %location of fixed bugs in the test program,

As a result, LeoFuzz found 23 previously unreported vulnerabilities, 11 of which are assigned CVE IDs and others have been confirmed by the corresponding developers. %, and seven of which received high severity score. 
Table \ref{tab-newbugs} presents the subject program, the buggy method, the type and CVE/Bug ID of each vulnerability.
%All these vulnerabilities have been confirmed by the corresponding developers, and 
Seven of eleven CVEs are assigned high severity score. Below we discuss one of them in detail to highlight the ability of LeoFuzz. %as an example. %\lhl{use multiple bugs to highlight the power of LeoFuzz.}

  \begin{figure}[t]
  \scriptsize
 \begin{lstlisting}[xleftmargin=2em,firstnumber=435]
 static int  swf_FontExtract_DefineTextCallback(int id, SWFFONT * f, TAG * t, int jobs, void (*callback) (void *self, int *chars, int *xpos, int nr, int fontid, int fontsize, int xstart, int ystart, RGBA * color), void *self)
 	\end{lstlisting}
     ...
 \begin{lstlisting}[xleftmargin=2em,firstnumber=494]
   for (i = 0; i < num; i++) {
     int glyph;
     int adv = 0;
     advance[i] = xpos;
     glyph = swf_GetBits(t, gbits);
     adv = swf_GetBits(t, abits);
 \end{lstlisting}
%     ...
 \caption{Code snippet of swftext.c in SWFTools project}
  	\label{fig:swftext}
  \end{figure}

  \begin{figure}[t]
  \scriptsize
  \begin{lstlisting}[xleftmargin=2em,firstnumber=1201]
 TAG * swf_ReadTag(reader_t*reader, TAG * prev)
  \end{lstlisting}
  ...
  	\begin{lstlisting}[xleftmargin=2em,firstnumber=1229]
   if (reader->read(reader, t->data, t->len) != t->len) {
     #ifdef DEBUG_RFXSWF
     fprintf(stderr, "rfxswf: Warning: Short read (tagid %d). File truncated?\n", t->id);
     #endif
     free(t->data);  t->data=0;
     free(t);
     return NULL;
   }
  	\end{lstlisting}
 \caption{Code snippet of rfxswf.c in SWFTools project}
  	\label{fig:rfxswf}
  \end{figure}
       
 LeoFuzz found a use-after-free vulnerability in SWFTools package, \ie CVE-2021-42203, which involves different functions in multiple files. As shown in Fig. \ref{fig:swftext} and Fig. \ref{fig:rfxswf}, the program uses a pointer $t$ of TAG type at the line \textit{swftext.c}:498, and frees it at the line \textit{rfxswf.c}:1234. When testing the program, LeoFuzz successfully explored a path where function \textit{swf\_ReadTag} is called before function \textit{swf\_FontExtract\_DefineTextCallback}, causing the program crash. %heap-use-after-free in function \texttt{swf\_FontExtract\_DefineTextCallback} of \texttt{swftext.c} 

\subsection{RQ6: Contributions of four design decisions}
To evaluate the contributions of four main design decisions in LeoFuzz, \ie target sequence enhancement, exploration-exploitation coordination, fine-grained energy scheduling strategy and concolic execution, we disabled each technique and compiled four variants of LeoFuzz and named them LeoFuzz-s, LeoFuzz-e, LeoFuzz-f and  LeoFuzz-c respectively. We ran LeoFuzz and its four variants against those programs in Table \ref{tab:put} and the results are shown in Table \ref{tab-RQ6}.
The TTE columns indict the mean value of TTEs spent by LeoFuzz-s, LeoFuzz-e, LeoFuzz-f, LeoFuzz-c and LeoFuzz to trigger all CVEs in each subject program, and the factor columns reflect the TTE's ratio between four variants and LeoFuzz.

We have two findings: 1) Each technique contributes to LeoFuzz as the performance of each variant is weaker than LeoFuzz; 2) The contribution of MES is better than that of CEE and that of target sequence enhancement, which are better than that of concolic execution. It is reasonable because MES considers both the relations between a seed and targets and the relations within multiple target sequences.

\begin{table*}
\renewcommand\arraystretch{1.2}
    \centering
    \caption{Effectiveness of each design decision in LeoFuzz}
    \label{tab-RQ6}
    \begin{tabular}{c|c|c|c|c|c|c|c|c|c}
    \hline
        \multirow{2}{*}{Program} &   \multicolumn{2}{c|}{LeoFuzz-s} &\multicolumn{2}{c|}{LeoFuzz-e} & \multicolumn{2}{c|}{LeoFuzz-f} & \multicolumn{2}{c|}{LeoFuzz-c} &  LeoFuzz \\
        \cline{2-10}
        &TTE & Factor &TTE & Factor & TTE  &Factor & TTE & Factor & TTE \\ \hline
        cxxfilt & 1h15m & 1.61 & 1h16m & 1.64 & 1h27m & 1.88 & 1h13m & 1.57  & 46m22s  \\
        objdump & --- & 2.69 & 7h51m & 1.76 & --- & 2.69 & 5h34m & 1.25 & 4h28m \\
        readelf & 2h17m & 2.45 & 3h7m & 3.40 & 3h41m & 3.95 & 2h18m & 2.46 & 56m11s\\
        tiff2pdf & 2h53m & 2.25 & 2h21m & 1.83 & 3h26m & 2.68 & 1h32m & 1.32 & 1h17m \\
        dwg2dxf & 4h23m & 1.14 & 5h23m & 1.40 & 6h19m & 1.64 & 4h36m & 1.20 & 3h51m  \\
        unzzipcat-mem & 1h49m & 1.06 & 2h16m & 1.32 & 2h31m & 1.47 & 1h53m & 1.10 & 1h43m  \\
        mjs & 17m52s & 1.35 & 25m10s & 1.92 & 39m14s & 3.00 & 17m41s & 1.31 & 13m27s  \\\hline
    \end{tabular}
\end{table*}

\section{Threats to validity} \label{threats}
%\cxl{We have identified the following threats to validity.}
%\begin{enumerate}
Internal validity: The main internal threat is the randomness of fuzzing.
We conducted the experiments multiple times for fairness, and as initial seeds might influence the outcomes in the fuzzing experiments, we used the same seeds to LeoFuzz as inputs to each baseline in all experiments. 
The second internal threat comes from the configurable options in LeoFuzz, \eg  two parameters in Equation \ref{eq1} and \ref{eq2}, which are currently set based on our preliminary experiments. Though the current results are promising, we believe fine-tuning
them may improve the experiment results but it is not the key technique here. Therefore, we leave it as future work.

External validity: Although our experimental results may vary to other programs, to mitigate this threat, we chose 31 vulnerabilities in 7 real-world programs that have been frequently evaluated in the existing fuzzers. These programs also have diverse functionalities as well as different program sizes. Moreover, the vulnerabilities chosen come from different types (9 in total) and thus have different difficulty to trigger them.
%we randomly select multiple programs from diverse application domains as benchmarks in our experiments, and randomly select 31 vulnerabilities with different difficulty to trigger them.

Construct validity: We compare different configurations of LeoFuzz according to the main techniques proposed in this paper, so we can understand that any effect on the results is due to their differences, and can also verify that the proposed strategies are all effective.
%\end{enumerate}

\section{Related work}\label{relwork}

\subsection{Coverage-based Greybox Fuzzing}  Greybox fuzzing is scalable and practical in finding bugs or vulnerabilities in software. AFL~\cite{afl} uses lightweight compile-time instrumentation, coverage feedback and genetic algorithm to generate test cases that can trigger vulnerabilities in programs. Compared with blackbox fuzzing~\cite{peach} and whitebox fuzzing~\cite{klee,angr}, greybox fuzzing has higher efficiency and effectiveness~\cite{afl,ossfuzz}.

To improve the exploration ability of greybox fuzzing,
Vuzzer~\cite{vuzzer} uses dynamic data-flow analysis in greybox fuzzing to maximize coverage and explore deeper paths. Angora~\cite{angora} solves path constraints by gradient descent algorithm to improve the coverage of branches. REDQUEEN~\cite{redqueen} leverages a lightweight input-to-state correspondence mechanism as an alternative to data-flow analysis and symbolic execution. GREYONE~\cite{GREYONE} exploits a data flow-sensitive fuzzing scheme since fuzzing based on traditional data flow analysis is inaccurate and slow. DeepFuzzer~\cite{deepfuzzer} first uses symbolic execution to generate qualified initial seeds that can help pass complex checks, then applies a statistical seed selection algorithm to balance mutation frequencies among different seeds. Its hybrid mutation strategy aims to balance global exploration and deep search.
To refine the seed scheduling of greybox fuzzing,
AFLFast~\cite{aflfast} shows that most test cases execute high-frequency paths, so AFLFast assigns more energy to those seeds which can pass through the low-frequency paths.
EcoFuzz~\cite{ecofuzz} proposes a variant of the Adversarial MultiArmed Bandit (VAMAB) model to model scheduling problems and balances exploration stage and exploitation stage for reasonable seed selection, while LeoFuzz coordinates two stages adaptively to balance the fuzzer's code coverage and directedness. %Ecofuzz aims to improve the performance of coverage-based fuzzing while LeoFuzz focuses on improving the target-oriented directed fuzzing.
AFLsmart~\cite{aflsmart} leverages a high-level structural representation of the seed file and new mutation operators to generate new files, and introduces a validity-based power schedule to spend more time generating files that are more likely to pass the parsing stage of the program.

%Like the above efforts, LeoFuzz is also designed to improve the fuzzing performance by proposing a fine-grained energy scheduling strategy and a novel approach to dynamically coordinate exploration and exploitation stages in fuzzing.
Unlike the above efforts that aim to improve the performance of coverage-based greybox fuzzing, LeoFuzz is a target-oriented directed greybox fuzzer that aims to trigger multiple target sites in a single instance.
PAFL~\cite{pafl} extends existing fuzzing optimizations of single mode to industrial parallel mode by dividing tasks and synchronizing guiding information. It improves the fuzzing efficiency by running more instances of multiple fuzzers and is complementary to our approach.

%aflgo, rdfuzz, hawkeye: seed distance guided.
%lolly, berry, uafl, sequence-guided
%savior, parmesan using sanitizer
%cafl, beacon, windranger newer 
\subsection{Directed Greybox Fuzzing} %Directed greybox fuzzing  tools usually use a distance-to-target guided approach. 
AFLGo~\cite{AFLGo} is the first directed greybox fuzzer (DGF), its simulated annealing-based power schedule gradually assigns more energy to seeds that are closer to the target sites while reduces energy for seeds that are far away.
Based on AFLGo, Hawkeye~\cite{hawkeye} supports indirect calls and adjusts its seed prioritization, power scheduling and mutation strategies adaptively to reach the target sites rapidly. However, Hawkeye has similar problems with AFLGo when dealing with multiple targets. As discussed in Section~\ref{motivation}, their unsuitable energy schedule may ignore local optimal solutions and hinder covering multiple targets efficiently. Moreover, their strategy of coordinating exploration-exploitation stages is static and inflexible. 
Also based on AFLGo, RDFuzz~\cite{RDFuzz} prioritizes a seed by combining its input-distance to the target sites and its trace's frequency and uses a \emph{static} intertwined schedule to perform exploration and exploitation in turn. By contrast, LeoFuzz dynamically coordinates exploration and exploitation stages according to the ratio of directed seeds and coverage seeds.

Some directed fuzzers exploit a sequence-based guided approach. Lolly~\cite{Lolly} is the first sequence directed greybox fuzzer. For a given set of target statement sequences, Lolly aims to generate inputs that can reach the statements in each sequence in order. Berry~\cite{Berry} uses the target program's CFGs to extend the given target sequence to improve the directedness of fuzzing. UAFL~\cite{uafl} focuses on UAF vulnerability and thus takes use-after-free sequence to guide its fuzzer.
Lolly, Berry and UAFL consider the execution order of targets, while LeoFuzz further takes into account three kinds of relation between seeds and targets, \ie seed's target sequence coverage, priority and global maximum coverage of each target sequence.

Several directed fuzzers leverage the output of sanitizers to guide fuzzing. SAVIOR~\cite{SAVIOR} uses the output of UBSan as target sites, and calculates a seed's energy according to the new branches that the seed meets, the target sites on these branches and the difficulty of solving these branches' constraints. ParmeSan~\cite{ParmeSan} leverages the errors or warnings reported by multiple sanitizers as target sites and then guides the fuzzer by the distance from a seed to a target.  LeoFuzz can also use the results from sanitizers as targets though it does not depend on sanitizers.

%In the last two years, some new directed fuzzing have been proposed.
%Recently several directed fuzzers appears in parallel with our work.
Several directed fuzzers use data flow analysis or data conditions.
CAFL~\cite{cafl} aims to satisfy a sequence of constraints and prioritizes the seeds that
better satisfy those in order. It defines a constraint as a single target site and optionally a number of data conditions. If multiple constraints are specified, they must be satisfied in the specified order. 
CAFL assumes that the target sites are dependent to each other while LeoFuzz support multiple independent target sites. CAFL requires the additional information sources, \ie  crash dumps from memory error detectors and changelogs from patches, to generate the constraints. Moreover, CAFL was evaluated with up to 2 targets and cannot cover bugs that require three or more constraints, while LeoFuzz triggered ten bugs in dwg2dxf within four hours in our evaluation. % \eg use-of-uninitialized-value and Call-stack-overflow.
Considering that each basic block isn't equally important in seed distance calculation,
WindRanger~\cite{windranger} uses the deviation basic blocks (DBBs) and their data flow information for seed distance calculation, mutation, seed prioritization and energy scheduling. It dynamically switches between the exploration and exploitation stage according to the execution status of DBBs. 
Beacon~\cite{beacon} leverages a provable path pruning method to improve the efficiency of DGF. It identifies infeasible paths via control flow reachability and path condition satisfiability, instruments those related statements, and prunes these paths during fuzzing.
By contrast, LeoFuzz combines concolic execution and fuzzing by sharing two types of seeds, and hence can solve the path constraint for a seed and mutate the validated inputs satisfying the path condition. 
Overall, these methods are orthogonal to LeoFuzz and can be integrated with LeoFuzz for better performance.

\section{Conclusion}\label{conclusion}
We present a multiple targets directed greybox fuzzing approach, which leverages a novel strategy to adaptively coordinate exploration and exploitation stages, and a novel energy scheduling strategy that considers more relations between seeds and targets, \ie target sequence coverage, target sequence priority and global maximum coverage of target sequence. Our approach also uses concolic execution to help the fuzzer explore complex branches in programs. We implement our approach in LeoFuzz and evaluate it on crash reproduction, true positives verification, and vulnerability exposure in seven real-world programs. Experimental results show that LeoFuzz outperforms six state-of-the-art tools, \ie QYSM, AFLGo, Lolly, Berry, Beacon and WindRanger.

As future work, we will combine parallel programming with LeoFuzz to improve its performance further. We are also planning to combine LeoFuzz with QEMU emulator to discover vulnerabilities in embedded devices.

\section*{Acknowledgments}
We would like to thank the anonymous reviewers for their insightful comments. 

% \section*{References}
% \newpage


\begin{thebibliography}{00}
\bibitem{b1}https://nvd.nist.gov/vuln/search/results?form\_type=Basic\&results\_type=\\overview\&query=objdump+2.34\&search\_type=all

\bibitem{b2} https://nvd.nist.gov/vuln/search/results?form\_type=Basic\&results\_type=\\overview\&query=readelf+2.28\&search\_type=all

\bibitem{cxxfilt}https://cve.mitre.org/cgi-bin/cvekey.cgi?keyword=binutils

\bibitem{b3}https://nvd.nist.gov/vuln/search/results?form\_type=Basic\&results\_type=\\overview\&query=tiff2pdf+4.0.9\&search\_type=all

\bibitem{b4}https://nvd.nist.gov/vuln/search/results?form\_type=Basic\&results\_type=\\overview\&query=tiff2pdf+4.0.8\&search\_type=all

\bibitem{b5}https://nvd.nist.gov/vuln/search/results?form\_type=Basic\&results\_type=\\overview\&query=jasper++2.0.14\&search\_type=all

\bibitem{b6}https://nvd.nist.gov/vuln/search/results?form\_type=Basic\&results\_type=\\overview\&query=jasper++2.0.12\&search\_type=all

\bibitem{b7}https://nvd.nist.gov/vuln/search/results?form\_type=Basic\&results\_type=\\overview\&query=tiff2pdf+4.0.9\&search\_type=all

\bibitem{b8} https://www.sqlite.org/cves.html

\bibitem{b9} https://nvd.nist.gov/vuln/search/results?form\_type=Basic\&results\_type=\\overview\&query=tcpdump+4.9.3\&search\_type=all

\bibitem{b10} https://httpd.apache.org/security/vulnerabilities\_24.html
\bibitem{fuzz}H. Liang, X. Pei, X. Jia, W. Shen, and J. Zhang, ``Fuzzing: State of the Art". IEEE Transactions on Reliability, vol.67, no.3 pp, 1199-1218, Sep. 2018
\bibitem{afl}M. Zalewski, ``American fuzzy lop," http://lcamtuf.coredump.cx/afl/.

\bibitem{AFLGo}M. Böhme, V.-T. Pham, M.-D. Nguyen, and A. Roychoudhury, ``Directed Greybox Fuzzing", in Proceedings of the 2017 ACM SIGSAC Conference on Computer and Communications Security, CCS 2017, Dallas, TX, USA, October 30 - November 03, 2017, 2017, pp. 2329–2344.
\bibitem{Lolly}H. Liang, Y. Zhang, Y. Yu, Z. Xie and L. Jiang, ``Sequence Coverage Directed Greybox Fuzzing," in Proceedings of 2019 IEEE/ACM 27th International Conference on Program Comprehension (ICPC), 2019, pp. 249-259.
\bibitem{Berry}H. Liang, L. Jiang, L. Ai and J. Wei, ``Sequence Directed Hybrid Fuzzing", in Proceedings of 2020 IEEE 27th International Conference on Software Analysis, Evolution and Reengineering (SANER), 2020, pp. 127-137.
\bibitem{QSYM}I. Yun, S. Lee, M. Xu, Y. Jang, and T. Kim, ``QSYM:A Practical
Concolic Execution Engine Tailored for Hybrid Fuzzing", in Proceedings of USENIX Security Symposium, 2018, pp. 745–761.
\bibitem{redqueen}C. Aschermann, S. Schumilo, T. Blazytko, R. Gawlik, and T. Holz, ``REDQUEEN: Fuzzing with Input-to-State Correspondence", in Proceedings of 2019 Network and Distributed System Security Symposium, 2019.
\bibitem{hawkeye}H. Chen, Y. Xue, Y. Li, B. Chen, X. Xie, X. Wu, and Y. Liu, ``Hawkeye: Towards a Desired Directed Grey-box Fuzzer", in Proceedings of the 2018 ACM SIGSAC Conference on Computer and Communications Security, CCS 2018, pp. 2095–2108.
\bibitem{SAVIOR}Y. Chen, P. Li, J. Xu, S. Guo, R. Zhou, Y. Zhang, T. Wei, and L. Lu, ``SAVIOR: Towards Bug-Driven Hybrid Testing", in Proceedings of 2020 IEEE Symposium on Security and Privacy (SP), pp. 1580–1596.
\bibitem{GREYONE}S. Gan, C. Zhang, P. Chen, B. Zhao, X. Qin, D. Wu and Z. Chen. "GREYONE: Data Flow Sensitive Fuzzing", in Proceedings of USENIX Security Symposium, 2020, pp. 2577-2594.
\bibitem{ParmeSan}S. Österlund, K. Razavi, H. Bos and C. Giuffrida, ``ParmeSan: Sanitizer-guided Greybox Fuzzing.", in Proceedings of USENIX Security Symposium, 2020, pp. 2289-2306.
\bibitem{uafl}H. Wang, X. Xie, Y. Li, C. Wen, Y. Li, Y. Liu, S. Qin, H. Chen and Y. Sui, ``Typestate-Guided Fuzzer for Discovering Use-after-Free Vulnerabilities.", in Proceedings of IEEE/ACM 42nd International Conference on Software Engineering (ICSE) ,2020, pp. 999-1010.
\bibitem{RDFuzz}J. Ye, R. Li and B. Zhang, ``RDFuzz: Accelerating Directed Fuzzing with Intertwined Schedule and Optimized Mutation.", in Proceedings of Mathematical Problems in Engineering, 2020, pp. 1-12.
\bibitem{aflfast}M. Böhme, V. Pham and A. Roychoudhury, ``Coverage-Based Greybox Fuzzing as Markov Chain," in IEEE Transactions on Software Engineering, vol. 45, no. 5, pp. 489-506.
\bibitem{vuzzer}S. Rawat, V. Jain, A. Kumar, L. Cojocar, C. Giuffrida, and H. Bos, ``VUzzer: Application-aware Evolutionary Fuzzing", In Proceedings 2017 Network and Distributed System Security Symposium, 2017.
\bibitem{klee}C. Cadar, D. Dunbar, and D. R. Engler, ``KLEE: Unassisted and Automatic Generation of High-Coverage Tests for Complex Systems Programs," in 8th USENIX Symposium on Operating Systems Design and Implementation, OSDI 2008, December 8-10, 2008, San Diego, California, USA, Proceedings, R. Draves and R. van Renesse, Eds. USENIX Association, 2008, pp. 209–224.
\bibitem{angr}Y. Shoshitaishvili et al., ``SOK: (State of) The Art of War: Offensive Techniques in Binary Analysis," 2016 IEEE Symposium on Security and Privacy (SP), 2016, pp. 138-157.
\bibitem{peach}Peach. 2016. Peach Fuzzer: Discover unknown vulnerabilities. Peach Fuzzer (July 2016). http://www.peachfuzzer.com/
\bibitem{ecofuzz} Yue, Tai, Pengfei Wang, Yong Tang, Enze Wang, Bo Yu, Kai Lu and Xu Zhou. “EcoFuzz: Adaptive Energy-Saving Greybox Fuzzing as a Variant of the Adversarial Multi-Armed Bandit.” USENIX Security Symposium,2020, pp.2307-2324.
\bibitem{cafl}L. Gwangmu, W.-J. Shim and B. Lee, ``Constraint-guided Directed Greybox Fuzzing.", in Proceedings of USENIX Security Symposium, 2021, pp. 3559-3576.
\bibitem{angora}P. Chen, H. Chen, ``Angora: Efficient Fuzzing by Principled Search," 2018 IEEE Symposium on Security and Privacy (SP), 2018, pp. 711-725.
\bibitem{fairfuzz}C. Lemieux and K. Sen, ``FairFuzz: Targeting Rare Branches to Rapidly Increase Greybox Fuzz Testing Coverage", arXiv:1709.07101 [cs], September 2017.

\bibitem{libfuzzer} libfuzzer, https://llvm.org/docs/LibFuzzer.html, 2015.
\bibitem{ossfuzz} Oss-fuzz report, https://security.googleblog.com/2018/11/a-new-chapter-for-oss-fuzz.html, 2018.

\bibitem{domtree}Finding Dominators in Directed Graphs. SIAM Journal on Computing. Society for Industrial and Applied Mathematics. Retrieved June 24, 2021 from https://epubs.siam.org/doi/10.1137/0203006

\bibitem{clang} Clang Static Analyzer. [Online]. Available: http://clanganalyzer.llvm.org/

\bibitem{binutils}Binutils is a binary toolset. [Online]. Available: https://github.com/bminor/binutils-gdb

\bibitem{libtiff}Libtiff is a library for reading and writing tag image file format (abbreviated as TIFF).[Online]. Available:http://www.libtiff.org/

\bibitem{libredwg}LibreDWG is a free C library to read and write DWG files.[Online]. Available: https://github.com/LibreDWG/libredwg

\bibitem{zziplib}zziplib offers the ability to easily extract data from files archived in a single zip file.[Online]. Available: http://zziplib.sourceforge.net/

\bibitem{mjs} mjs is designed for microcontrollers with limited resources. [Online]. Available: http://github.com/cesanta/mjs

\bibitem{libming}Libming is a library for generating macromedia flash files. [Online]. Available: https://github.com/libming/libming

\bibitem{swftools} SWFTools is a collection of utilities for working with Adobe Flash files (SWF files). [Online]. Available: http://www.swftools.org/

\bibitem{clusterfuzz}ClusterFuzz is a scalable fuzzing infrastructure that finds security and stability issues in software. [Online]. Available: https://github.com/google/clusterfuzz

\bibitem{pafl}Jie Liang, Yu Jiang, Yuanliang Chen, Mingzhe Wang, Chijin Zhou and Jiaguang Sun. ``PAFL: extend fuzzing optimizations of single mode to industrial parallel mode." Proceedings of the 26th ACM Joint Meeting on European Software Engineering Conference and Symposium on the Foundations of Software Engineering, 2018. pp. 809--814.
\bibitem{aflsmart}Van-Thuan Pham, Marcel Böhme, Andrew E. Santosa, Alexandru Razvan Caciulescu, Abhik Roychoudhury. ``Smart Greybox Fuzzing." IEEE Trans. Software Engineering. 47(9): 1980-1997, 2021.
\bibitem{deepfuzzer}Jie Liang, Yu Jiang, Mingzhe Wang, Xun Jiao, Yuanliang Chen, Houbing Song, Kim-Kwang Raymond Choo. ``DeepFuzzer: Accelerated Deep Greybox Fuzzing." IEEE Trans. Dependable Secure Computing. 18(6): 2675-2688, 2021.
\bibitem{windranger}Zhengjie Du, Yuekang Li, Yang Liu, and Bing Mao. ``WindRanger: A Directed Greybox Fuzzer driven by Deviation Basic Block." 44th International Conference on Software Engineering (ICSE), 2022.
\bibitem{beacon}Heqing Huang, Yiyuan Guo, Qingkai Shi, Peisen Yao, Rongxin Wu and Charles Zhang. “BEACON: Directed Grey-Box Fuzzing with Provable Path Pruning.”, The 43rd IEEE Symposium on Security and Privacy. 2022.
\bibitem{a12} A. Vargha and H. D. Delaney, “A Critique and Improvement of the Common Language Effect Size Statistics of McGraw and Wong,” Journal of Educational \& Behavioral Statistics, vol. 25, no. 2, pp. 101–132, 2000.



\end{thebibliography}
\end{document}